\documentclass[camera,letterpaper,nomarginnotes,nonarrowgutter]{jpaper}
\newif\ifdraft
\draftfalse

\usepackage{tikz}
\usepackage{mathptmx} 
\usepackage{amsmath,amssymb,amsfonts}
\usepackage{comment}
\usepackage{graphics}
\usepackage{fancyhdr}
\usepackage{booktabs}
\usepackage{pifont}

\usepackage[normalem]{ulem}
\usepackage{cite}
\usepackage{hhline}
\usepackage{xcolor}
\usepackage{setspace}
\usepackage{textcomp}
\usepackage{float}
\usepackage{enumitem}
\usepackage{afterpage}
\usepackage{graphicx}
\usepackage[htt]{hyphenat}
\usepackage{makecell}
\usepackage{xspace}
\usepackage{listings}
\usepackage{multirow}
\usepackage{balance}
\usepackage{glossaries} 
\usepackage{siunitx}
\usepackage{duckuments} 
\usepackage{dblfloatfix} 
\usepackage{subcaption}
\usepackage{rotating}
\usepackage[us,12hr]{datetime}
\usepackage[en-GB, useregional=numeric]{datetime2}

\usepackage[compact]{titlesec}
\usepackage[colorinlistoftodos,prependcaption,textsize=small]{todonotes} 
\usepackage{setspace}
\usepackage{dblfloatfix} 
\usepackage{algorithm2e}
\usepackage{soul}
\usepackage{bm}

\usepackage[bookmarks=true,breaklinks=true,hidelinks]{hyperref}

\newif\ifarxiv 
\arxivtrue

\newif\ifcameraready
\camerareadyfalse

\newif\ifrev
\revfalse

\newif\ifpagenumbers
\pagenumberstrue

\newcounter{version}
\ifcameraready
\else
\fi

\newcommand\proposal{Ariadne\xspace}
\newcommand\dataorg{HotnessOrg\xspace}
\newcommand\compress{AdaptiveComp\xspace}
\newcommand\predi{PreDecomp\xspace}

\title{\Large Ariadne: A Hotness-Aware and Size-Adaptive Compressed Swap Technique \\  for Fast Application Relaunch and Reduced CPU Usage on Mobile Devices \vspace{-0.6em}}
\newcommand{\affilETH}[0]{\textsuperscript{$\S$}}
\newcommand{\affilCITYU}[0]{\textsuperscript{$\dagger$}}
\newcommand{\affilMBZUAI}[0]{\textsuperscript{$^\ddagger$}}
\newcommand{\affilKCL}[0]{\textsuperscript{$^\P$}}
\newcommand{\affilUST}[0]{\textsuperscript{$\mathcal{S}$}}
\newcommand{\affilMangoBoost}[0]{\textsuperscript{$^\star$}}

\author{
{Yu Liang\affilETH}
{Aofeng Shen\affilETH}
{Chun Jason Xue\affilMBZUAI}
{Riwei Pan\affilCITYU}
{Haiyu Mao\affilKCL}
{Nika Mansouri Ghiasi\affilETH}\\
{Qingcai Jiang\affilUST}
{Rakesh Nadig\affilETH}
{Lei Li\affilCITYU}
{Rachata Ausavarungnirun\thanks{This work was done before the author joined MangoBoost.} \affilMangoBoost}
{Mohammad Sadrosadati\affilETH}
{Onur Mutlu\affilETH} \vspace{0.4em}\\
{\small \affilETH \emph{ETH Z{\"u}rich} \quad
\affilMBZUAI \emph{MBZUAI}\quad
\affilCITYU \emph{City University of Hong Kong} \quad
\affilKCL \emph{King's College London}} \\ 
{\small \affilUST \emph{University of Science and Technology of China}\quad
\affilMangoBoost \emph{MangoBoost}}
}

\ifarxiv
    \pagenumberstrue
\else
    \ifcameraready
    \else 
        \newcounter{passversion}
        \pagenumberstrue
        \fancypagestyle{firstpage}
        {
            \fancyhead{}
            \fancyhead[C]{\textcolor{red}{CONFIDENTIAL DRAFT -- DO NOT DISTRIBUTE -- TO APPEAR IN DRAMSec'24} \\ \textcolor{MidnightBlue}{\emph{Version \thepassversion{}.4~---~\today, \ampmtime}} }
        }
    \fi
\fi

\makeatletter
\def\bstctlcite{\@ifnextchar[{\@bstctlcite}{\@bstctlcite[@auxout]}}
\def\@bstctlcite[#1]#2{\@bsphack
  \@for\@citeb:=#2\do{%
    \edef\@citeb{\expandafter\@firstofone\@citeb}%
    \if@filesw\immediate\write\csname #1\endcsname{\string\citation{\@citeb}}\fi}%
  \@esphack}
\makeatother

\begin{document}
\bstctlcite{IEEEexample:BSTcontrol}
\maketitle
\ifcameraready
    \thispagestyle{empty}
    \pagestyle{empty}
\else 
    \pagestyle{plain}
\fi

\ifcameraready
    \setcounter{version}{99}
\else
    \setcounter{version}{99}
\fi

\begin{abstract}

As the memory demands of individual mobile applications continue to grow and the number of concurrently running applications increases, available memory on mobile devices is becoming increasingly scarce. 
When memory pressure is high, current mobile systems use a RAM-based compressed swap scheme (called ZRAM) to compress unused execution-related data (called anonymous data in Linux) in main memory. 
This approach avoids swapping data to secondary storage (NAND flash memory) or terminating applications, thereby achieving shorter application relaunch latency.

In this paper, we observe that the state-of-the-art ZRAM scheme prolongs relaunch latency and wastes CPU time because it does \emph{not} differentiate between hot and cold data or leverage different compression chunk sizes and data locality.
We make three new observations. First, anonymous data has different levels of hotness. Hot data, used during application relaunch, is usually similar between consecutive relaunches.
Second, when compressing the same amount of anonymous data, small-size compression is very fast, while large-size compression achieves a better compression ratio.
Third, there is locality in data access during application relaunch.

Based on these observations, we propose a hotness-aware and size-adaptive compressed swap scheme, \proposal, for mobile devices to mitigate relaunch latency and reduce CPU usage. \proposal incorporates three key techniques.
First, a low-overhead hotness-aware data organization scheme aims to quickly identify the hotness of anonymous data without significant overhead.
Second, a size-adaptive compression scheme uses different compression chunk sizes based on the data's hotness level to ensure fast decompression of hot and warm data. 
Third, a proactive decompression scheme predicts the next set of data to be used and decompresses it in advance, reducing the impact of data swapping back into main memory during application relaunch.
 
We implement and evaluate \proposal on a commercial smartphone,
Google Pixel 7 with the latest Android 14.
Our experimental evaluation results show that, on average, \proposal reduces application relaunch latency by 50\% and decreases the CPU usage of compression and decompression procedures by 15\% compared to the state-of-the-art compressed swap scheme for mobile devices.

\end{abstract}
\section{Introduction}
\label{sec:Intro}
Mobile devices are integral to our daily lives, with users frequently relaunching and running various applications to meet their diverse needs~\cite{lebeck2020end, low-end2, deng2019measuring}. To fulfill user expectations of seamless and rapid application relaunch, mobile systems preserve all execution-related data (called \emph{anonymous data} in Linux~\cite{anonymous}), such as stack and heap, in main memory. This practice, known as \emph{keeping applications alive in the background}~\cite{kim2019ezswap, bergman2022znswap, lebeck2020end, son2021asap, end2024more},
enables faster relaunches. However, it also results in significant main memory capacity requirements for each application. 

As the demand for memory capacity in mobile applications grows and the number of applications running simultaneously increases,  available memory is becoming an increasingly scarce resource on mobile devices~\cite{liang2020acclaim, son2021asap, end2024more, lebeck2020end}.
When memory capacity pressure is high, current mobile systems use a \emph{RAM-based compressed swap scheme} (called \emph{ZRAM}~\cite{zram1,zram2}) to compress unused anonymous data into a specific memory region, called \emph{zpool}~\cite{zpool}, rather than directly swapping the data into secondary storage (i.e., NAND flash memory). This approach achieves shorter application relaunch latency because decompression is much faster than swapping data from secondary storage into main memory and relaunching terminated applications.

We observe that the state-of-the-art \emph{ZRAM} scheme still prolongs relaunch latency and wastes CPU time due to two major reasons. First, it does not differentiate between hot and cold data, resulting in frequent and unnecessary compression and decompression. Systems might compress hot data when an application is in the background and decompress it when brought back to the foreground, even though enough memory space is available. The unnecessary compression and decompression not only prolongs application relaunch times but also wastes CPU resources. Second, it does not take advantage of different compression chunk sizes and data locality, leading to long compression and decompression times during application relaunch. As users typically switch between applications more than 100 times daily~\cite{deng2019measuring}, these frequent long-latency relaunches can negatively impact the overall user experience on mobile devices~\cite{predict1, predict2, predict3, context-aware, ubcomp6, ubcomp21, ubcomp23, ubcomp34, context, predict-privacy1, predict-privacy2, predict-privacy3, mobisysfastapp}.

The \textbf{goal} of this work is to minimize application relaunch latency and reduce wasted CPU  usage while maximizing the number of live background applications for an enhanced user experience.
To achieve this goal, we characterize the anonymous data of real mobile applications used on a real modern mobile phone (i.e., Google Pixel 7~\cite{Pixel7}). Our experimental characterization yields three new observations. First, we classify anonymous data into three categories: i) Hot data, used during relaunch and impacting relaunch latency; ii) Warm data, potentially used during application execution after relaunch; and iii) Cold data, usually not used again. We observe that hot data is usually similar between consecutive relaunches.
Second, when compressing the same amount of anonymous data, small-size compression, which involves compressing data in smaller chunks, is very fast, while large-size compression achieves a better compression ratio.
Third, there is locality in data access in \emph{zpool} when swapping in anonymous data during application relaunch, meaning the data tends to be stored in contiguous or nearby memory locations in \emph{zpool}. Thus, we can predict the next set of data to be used at the beginning of a relaunch.

Based on these observations, we propose a new hotness-aware and size-adaptive compressed swap scheme for mobile devices, called \proposal, that incorporates \textbf{three key techniques}:
First, a low-overhead hotness-aware data organization scheme aims to separate hot and cold data. \proposal tries to maintain hot data and compressed warm data in main memory while swapping compressed cold data to secondary storage.
Second, a size-adaptive compression scheme takes advantage of different compression chunk sizes. It uses small-size compression for identified hot and warm data to ensure fast relaunch and execution, while using large-size compression for cold data to achieve a high compression ratio.
Third, a proactive decompression scheme predicts the next set of data to be used and performs decompression in advance for such data, further mitigating the negative impact of data swapping back into main memory and decompression latency on application relaunch.

We implement and evaluate \proposal on a commercial smartphone,
Google Pixel 7~\cite{Pixel7} with the latest Android 14 operating system~\cite{Android14}.
We test \proposal with over 30 combinations of commonly-used concurrently-running mobile applications. Our experimental evaluation results show that, on average, \proposal reduces application relaunch latency by 50\% and decreases the CPU usage of compression and decompression procedures by 15\% compared to the state-of-the-art compressed swap scheme for mobile devices.

This work makes the following key contributions:
\begin{itemize}
    \item We are the first to quantitatively demonstrate the inefficiency of the state-of-the-art compressed swap scheme in mobile systems, highlighting its long application relaunch latency and high CPU usage, and identifying the root causes of these problems.  
    \item We make three new observations from real mobile applications. First, data used during application relaunch is usually similar between consecutive relaunches. Second, when compressing the same amount of anonymous data, small-size compression is very fast, while large-size compression achieves a better compression ratio. Third, there is locality in data access when swapping in anonymous data during application relaunch.
    \item We propose a new hotness-aware and size-adaptive compressed swap scheme, \proposal, for mobile devices. This scheme incorporates three key techniques: low-overhead hotness identification, size-adaptive compression, and proactive and predictive decompression.
    \item We evaluate \proposal on a real smartphone with a cutting-edge Android operating system. Our evaluation results show that our solution surpasses the state-of-the-art in terms of both application relaunch latency and CPU usage. To foster further research in the design and optimization of mobile compressed swap techniques, we open-source our implementations at https://github.com/CMU-SAFARI/Ariadne.
\end{itemize}

\section{Background and Motivation}
\label{sec:back}
Mobile devices have unique features such as fewer foreground applications, smaller DRAM and flash memory capacity, and constrained power budget compared to general-purpose servers. However, mobile systems, especially Android~\cite{Android}, are built on the Linux kernel~\cite{Linuxkernel}, originally designed for servers. As a result, many existing schemes in Android systems~\cite{liang2020acclaim, TC, son2021asap, lebeck2020end} do not align well with mobile workloads, leading to suboptimal performance, reduced device lifespan~\cite{liang2022cachesifter, liu2017non, zhu2017smartswap, guo2015mars}, and increased energy consumption~\cite{googleworkloads}. This section highlights the detrimental effects of the inefficient  \emph{ZRAM} scheme~\cite{zram1,zram2} on Android systems, especially focusing on application relaunch latency and energy efficiency.

\subsection{Application Launching and Execution}
We first briefly describe the mobile application launching procedure and then explain how to keep an application alive in mobile systems.

\noindent\textbf{Mobile application launching.}
Application launch latency is one of the critical metrics used to evaluate the user experience on mobile devices~\cite{predict1,predict2,predict3, context-aware, son2021asap}. It directly reflects the system's responsiveness and smoothness, as faster launch times contribute to a more immediate and seamless user experience. There are two types of application launching: ~\textit{cold launch} and ~\textit{hot launch}.
Cold launching an application involves two main steps: i) creating one or more processes for the application, and ii) reading the application's data into main memory. In contrast, a hot launch means launching an application from the background, so it does not require process creation as the application's processes are already running in the background. Previous work~\cite{liu2017non} shows that process creation accounts for 94\% of the total cold launch latency. Many studies~\cite{lebeck2020end, son2021asap, end2024more} show that hot launch is much faster than cold launch, leading to an improved user experience. Keeping applications alive in the background enables hot launches for relaunching applications, thereby reducing relaunch latency.

\noindent\textbf{Keeping mobile applications alive.} 
To determine how to keep an application alive, we first analyze its execution. Mobile application execution typically generates two types of memory pages: \emph{file-backed pages} and \emph{anonymous pages}. 
\emph{File-backed pages} directly correspond to files stored in secondary storage (e.g., NAND flash memory). When the system encounters insufficient available main memory, it frees up (i.e., reclaims)  memory pages to accommodate new requests. The system writes data from a reclaimed \emph{file-backed page} back to secondary storage. In contrast, \emph{anonymous pages} do \emph{not} correspond to any specific file in secondary storage but contain data associated with process execution, such as stack and heap information (called \emph{anonymous data}). When the system reclaims an \emph{anonymous page}, it deletes the \emph{anonymous data}, leading to the termination of the corresponding process. Therefore, to keep an application alive, it is essential to keep its \emph{anonymous data} in main memory.
To assess the feasibility of keeping a large number of applications alive, we measured the \emph{anonymous data} volumes of five commonly-used applications on a Google Pixel 7 (see experimental setup details in Section~\ref{sec:evaluation}).
We gathered data for each application at two time points: 10 seconds and 5 minutes after launching. The results are presented in Table~\ref{tab:anonymous}, which leads to two main observations.

\begin{table}[h!]
\vspace{0.3em}
\caption{Anonymous data volume (in MB) of five applications, where `GEarth' refers to Google Earth.}
\label{tab:anonymous}
\centering
\footnotesize
\begin{tabular}{c||c|c|c|c|c}
\hline
\textbf{Time}&\textbf{Youtube} & \textbf{Twitter}& \textbf{Firefox} & \textbf{GEarth}&\textbf{BangDream}\\ \hline
\hline
10s&177&182&560&273&326\\\hline
5mins&358&273&716&429&821\\\hline
\end{tabular}
\end{table}

First, each application generates substantial anonymous data, reaching up to 821 MB. Second, the volume of anonymous data increases as the application continues to run.
We conclude that each application generates a significant amount of anonymous data during its execution. Consequently, mobile systems require a substantial amount of main memory to keep all applications alive. However, due to cost and power constraints, main memory capacity in mobile devices is typically limited, ranging from 1 GB to 8 GB in low/mid-end smartphones~\cite{li2023ice}.  
The available memory on these smartphones for applications is usually limited and allows only a moderate number of applications to run concurrently in the background~\cite{end2024more,lebeck2020end}.

\subsection{Android Memory Swap Schemes}
To keep more applications alive on mobile devices with limited DRAM capacity, \emph{flash memory-based swap schemes} are employed to expand available memory space by relocating inactive \emph{anonymous pages} to a specific region in flash memory storage (i.e., flash memory swap space) ~\cite{zhu2017revisiting, kim2015cause, liu2017non, zhong2017building, kim2019analysis, zhu2017smartswap, kim2018comparison, oliveira2021extending, oliveira2021extending1}. These \emph{flash memory-based swap schemes} have two main issues. First, they increase the number of writes to flash memory storage, accelerating the wear-out of flash cells~\cite{cai2017error} and consequently reducing the overall lifespan of the mobile device~\cite{luo2015warm, liang2022cachesifter, liu2017non}. Second, compared to reading relaunch data directly from main memory, swapping data from flash memory storage into main memory can increase application relaunch latency~\cite{zhu2017smartswap,end2024more}, negatively impacting user experience.

\begin{figure}[!h]
\centering
\includegraphics[width=0.475\textwidth]{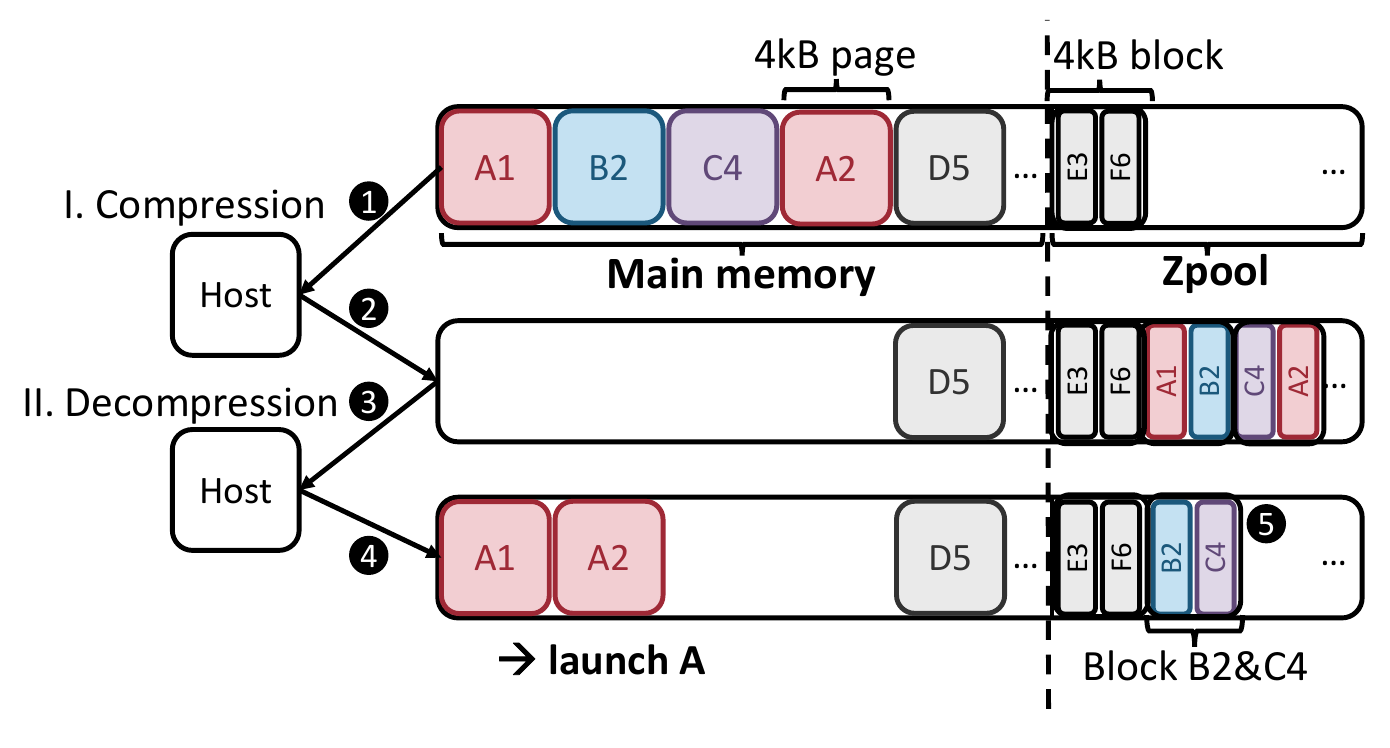}
\vspace{-0.4em}
\caption{Data movement flow for compression and decompression when using  \emph{ZRAM} in Android systems. Ai (e.g., A1, A2) represents an uncompressed anonymous page of an application. Block B2\&C4 refers to a compressed block that includes the compressed data of pages B2 and C4.}
\label{fig:datamove}
\end{figure}

\noindent\textbf{Cutting-edge compressed swap schemes.} 
To address the issues of \emph{flash memory-based swap schemes}, cutting-edge mobile devices often use a RAM-based compressed swap scheme (called \emph{ZRAM})~\cite{compress1, compress2, zram1, zram2, new-zram, merge-zram,oliveira2021extending, oliveira2021extending1}. 
Under \emph{ZRAM}, the system compresses unused anonymous data and stores it in a dedicated region of DRAM called \emph{zpool}. When the system relaunches an application, it decompresses the corresponding data from \emph{zpool}~\cite{zpool} back into main memory to facilitate the application's relaunch.
Figure~\ref{fig:datamove} illustrates the data movement for compression and decompression when utilizing  \emph{ZRAM}. 

Multiple applications (i.e., A, B, C, D, E, and F) concurrently run and continuously generate anonymous pages.  When available memory becomes insufficient, the system identifies and moves a set of least-recently-used (LRU~\cite{LRU}) data pages (pages A1, B2, C4, and A2 in Figure~\ref{fig:datamove}, where different letters represent different applications) to the host CPU or accelerator for compression \ding{182}.  The system stores the compressed data in 4KB memory blocks~\cite{ZRAMunit}. 
Next, it writes the compressed data (blocks A1\&B2 and C4\&A2) back to \emph{zpool} in DRAM \ding{183}. 
When a user launches an application (e.g., A), the system reads the application A-related compressed blocks (blocks A1\&B2 and C4\&A2) from  \emph{zpool} to the host CPU \ding{184} and decompresses them. 
The system writes the decompressed pages A1 and A2 back to main memory to facilitate application A's relaunch \ding{185}. Finally, the system merges the unused compressed data and writes block B2\&C4 back to  \emph{zpool} \ding{186}. 
As a result, the compression and decompression procedures due to \emph{ZRAM} can incur high-cost data movement, as quantified in~\cite{googleworkloads}.

\noindent\textbf{ \emph{ZRAM} with writeback (\emph{ZSWAP}).} 
With  \emph{ZRAM}, when  \emph{zpool} space is insufficient, the system deletes some inactive compressed data, potentially leading to application termination. \emph{ZSWAP} extends  \emph{ZRAM} by using flash memory storage for additional swapping space. When  \emph{zpool} is full, the system writes some compressed data to the flash memory-based swap space. While \emph{ZSWAP} increases the number of live background applications, it can also prolong application relaunch latency since some data needs to be swapped in from flash memory and decompressed during application relaunch. A major industrial vendor~\cite{oppo} reports that simply enabling \emph{ZSWAP} could lead to a 6x increase in application relaunch latency.  Due to such long relaunch latencies,  multiple vendors (e.g., Google~\cite{Pixel5, Pixel7} and Samsung~\cite{Galaxys}) may \emph{not} enable \emph{ZSWAP}. Therefore, the state-of-the-art compressed swap scheme is \emph{ZRAM} shown in Figure~\ref{fig:datamove}.

\subsection{Motivation}
\label{subsec:motivation}

We describe the major issues with the state-of-the-art compressed swap scheme,  \emph{ZRAM}, used in modern mobile devices.

\begin{figure}[!h]
\centering
\includegraphics[width=0.47\textwidth]{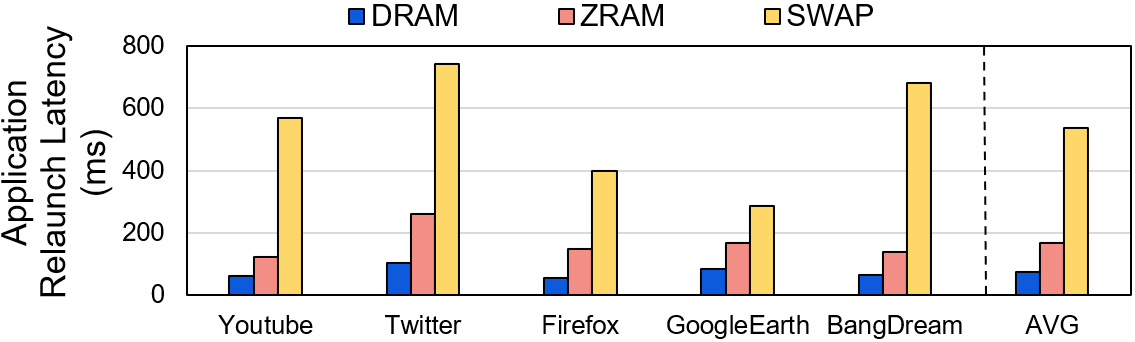}
\caption{Application relaunch latency under different memory swap schemes.}
\label{fig:relaunchlatency}
\end{figure}

\noindent\textbf{ \emph{ZRAM's} impact on performance.}  
To quantitatively demonstrate the impact of  \emph{ZRAM} on the performance of commercial mobile devices, we evaluate the hot launch (i.e., relaunch) latency. We choose application relaunch latency for two reasons. First,  \emph{ZRAM} directly impacts application relaunch latency since it needs to decompress the anonymous pages required for application relaunch. Second, relaunches occur frequently in users' daily lives (more than 100 times per day~\cite{deng2019measuring}), making it a crucial metric for evaluating performance~\cite{son2021asap}. Users typically perceive system responses as instantaneous if they occur within 100 ms~\cite{response}.
Figure~\ref{fig:relaunchlatency} shows the application relaunch latency of five commonly-used applications under three different swap schemes on Google Pixel 7: 1)  \emph{DRAM}, where the system reads all application data directly from DRAM (with the optimistic assumption that DRAM is large enough to host all such data), i.e., there is no swapping. 2) \emph{ZRAM}, where i) some application data is read directly from DRAM as it is stored uncompressed; ii) most application data is read from DRAM after being decompressed as it is stored in compressed form in \emph{zpool}.
Decompression may also trigger on-demand compression operations when main memory is insufficient, as the system must first compress other data to free up space for the decompressed data in main memory. 3) \emph{SWAP}, where the system reads data from the flash memory-based swap space into main memory during a relaunch. It does not involve compression or decompression. 

Based on the results, we observe that  \emph{ZRAM} outperforms the flash memory-based \emph{SWAP} scheme, but compression and decompression latencies still prolong application relaunch latency by an average of $2.1\times$ compared to reading data directly from DRAM without any compression or decompression. 

\noindent\textbf{Observation 1:} \textit{The state-of-the-art compressed swap scheme for mobile devices can lead to long application relaunch latencies due to long latencies of on-demand compression and decompression.} 

To avoid these latencies, some vendors, such as Google~\cite{Galaxys}, aggressively free up memory by proactively and periodically compressing data~\cite{Pixel7}. While this approach reduces the frequency of on-demand compression and decompression in \emph{ZRAM}, as shown in Figure~\ref{fig:relaunchlatency}, it also increases CPU usage.

\begin{figure}[!h]
\centering
\includegraphics[width=0.34\textwidth]{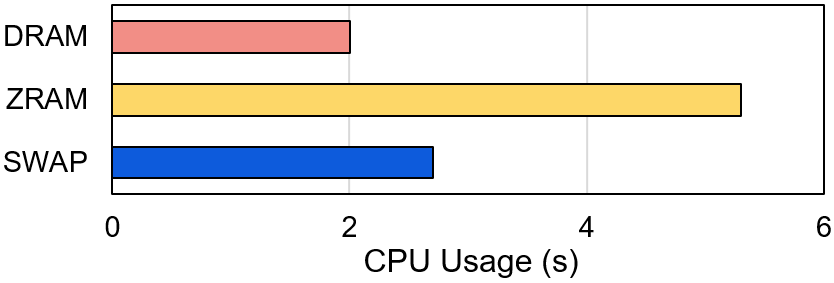}
\caption{CPU usage of the memory reclamation procedure (i.e., \emph{kswapd}) across different swap schemes.}
\label{fig:cputime}
\end{figure}

\noindent\textbf{\emph{ZRAM's} impact on CPU usage.}
To demonstrate the impact of  \emph{ZRAM} on CPU usage, we evaluate the CPU usage of the memory reclaim procedure under different memory swap schemes. 
We run the same application combinations as Figure~\ref{fig:relaunchlatency} to trigger memory swapping for a total of 60 seconds under different swap schemes (details in Section~\ref{sec:evaluation}) and use the system profiling tool, Perfetto~\cite{Perfetto}, to collect the CPU usage of the memory reclamation thread.
We test each swap scheme five times and calculate the average CPU time for each.
Figure~\ref{fig:cputime} shows the CPU usage of the reclamation thread (i.e., \emph{kswapd} thread) across different swap schemes on Google Pixel 7: 1) DRAM, where there is no swap scheme for \emph{anonymous data}, so the CPU usage includes CPU time used for writing \emph{file-backed} pages back to flash memory. 2)  \emph{ZRAM}, where the results also include the CPU usage for compressing \emph{anonymous data}.  However, the decompression procedure is not included because Perfetto can only track CPU usage for dedicated threads (e.g., reclaim thread), and decompression is not handled by such a thread. Therefore, ZRAM’s CPU usage is actually higher than what we report here. 3) \emph{SWAP}, where the CPU usage is collected while the system reclaims \emph{anonymous data} by writing it to flash
memory (Such usage can be low because when data is written to the storage device, CPU is usually yielded to other processes).

Based on the results, we observe that  \emph{ZRAM} increases CPU usage of memory reclamation by an average of $2.6\times$ compared to \emph{DRAM} and $2.0\times$ compared to \emph{SWAP}. 
Although the CPU usage of memory reclamation accounts for only a small percentage of the total CPU usage in the test scenario, it could become severe under heavy workloads, making it critical to reduce CPU usage for mobile devices~\cite{zhong2015energy, hort2021survey, chang2019lsim, nguyen2013storage}. Notably, the memory footprint of applications and systems is expected to grow in the future, e.g., with the rise of emerging generative artificial intelligence (GenAI) models~\cite{yin2024llm, karapantelakis2024generative, wen2023empowering} and augmented reality (AR) games~\cite{AR1, AR2, AR3, AR5}. As a result, \emph{ZRAM}'s CPU usage for compression and decompression is anticipated to increase, as memory capacity will remain constrained by cost and power limitations.

We also evaluate the total energy consumption of Google Pixel 7 under the above three swap schemes across two usage scenarios: light workloads (switching between ten applications (described in Section~\ref{sec:evaluation}) with 1-second intermission time in between) and heavy workloads (launching ten applications sequentially without any intermission time). We collect energy consumption for 60 seconds using Power Rails~\cite{power-profiler}. The test is repeated five times, and we report the average results in Table~\ref{tab:energy}. The results show that  \emph{ZRAM} increases energy consumption by 12.2\% under light workloads and 19.5\% under heavy workloads, compared to \emph{DRAM}.
 
\begin{table}[h!]
\centering
\caption{Energy consumption under three swap schemes.}
\footnotesize
\begin{tabular}{c||c|c|c|c}
\hline
\textbf{Workload} & & \textbf{DRAM}& \textbf{ZRAM}& \textbf{SWAP}\\ \hline
\hline
Light &Energy (J)&178.8 &200.7&179.4\\
&Normalized&1.000&1.122&1.003\\
\hline
Heavy &Energy (J)&231.8 &277.0&235.8\\
&Normalized& 1.000&1.195&1.017\\
\hline
\end{tabular}
\label{tab:energy}
\end{table}

\begin{figure*}[!h]
\centering
\includegraphics[width=0.98\textwidth]{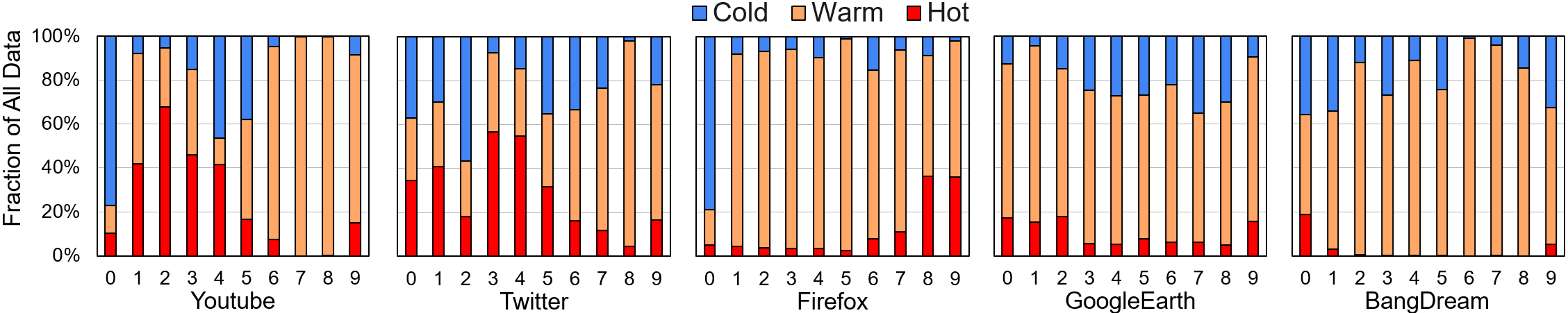}
\caption{Proportion of hot, warm, and cold data in each part of compressed data. We sort all compressed data in the order of compression time and then divide it into ten equal parts (X-axis). The data in part 0 is the first to be compressed, that in part 8 is the last.}
\label{fig:coldwarmhot}
\end{figure*}

\noindent\textbf{Observation 2:} \textit{The state-of-the-art compressed swap scheme for mobile devices consumes significant CPU time and energy.}

\noindent\textbf{Analysis of data compressed by \emph{ZRAM}.}
To determine the root causes of long relaunch latency and high CPU usage caused by  \emph{ZRAM}, we profile the swapped data. 
We analyze the swap process by sorting all compressed data in the order of compression time and then dividing them into ten equal-sized parts.
The data in part 0 is compressed first. To minimize swapping, cold data \emph{should} be swapped out (i.e., compressed) earlier (e.g., in parts 0 and 1) and hot data later (e.g., in parts 8 and 9).

Figure~\ref{fig:coldwarmhot} shows the proportion of hot, warm, and cold data in each part. 
The results indicate that mobile systems do \emph{not} consider the hotness of the data when swapping/compressing data. For example, the first part (i.e., part 0) of the swapped data includes a significant amount of hot data for \emph{all} applications. This is because the system still relies on the LRU scheme ~\cite{LRU} for choosing which data to swap, even though the LRU scheme may \emph{not} be useful for performance, especially when applications are switched often.

\noindent\textbf{Observation 3:} \textit{Compressing data without distinguishing between hot and cold data is the primary cause of long relaunch latency and high CPU usage, as it leads to frequent compression and decompression.} 

\noindent\textbf{Summary.} We empirically observe that the state-of-the-art  \emph{ZRAM} scheme prolongs application relaunch latency and consumes substantial CPU time because it does \emph{not} consider hotness when compressing data, leading to unnecessary compression and decompression operations.
There are numerous previous works~\cite{son2021asap, bergman2022znswap, Changlong2020seal, end2024more} that focus on optimizing flash memory-based swap schemes for mobile devices. However, most modern Android systems (e.g.,~\cite{Pixel7, Pixel5, Galaxys}) adopt the \emph{ZRAM} scheme~\cite{new-zram, zram1, zram2} instead of flash memory-based swap schemes due to its better performance. No existing work specifically aims to reduce relaunch time and CPU usage by optimizing the compressed swap scheme on mobile devices. 

\subsection{Our Goal}

\textbf{Our goal} in this work is to design a new compressed swap scheme for mobile devices that minimizes application relaunch latency and CPU usage while maximizing the number of live applications for enhanced user experience. Doing so requires reducing the frequency and latency of compression and decompression. We anticipate two \textbf{challenges} to achieving our goal. First, we would like to minimize compression and decompression frequency while maintaining efficient memory utilization. Second, we would like to reduce compression and decompression latency without negatively impacting the compression ratio.

To address these challenges, we profile mobile workloads with the goal of identifying new opportunities for designing a more efficient compressed swap scheme for mobile devices.

\section{New Insights into Mobile Workloads} 
\label{sec:insights}
We profile the \emph{anonymous data} of mobile workloads on the Google Pixel 7 (see Section~\ref{sec:evaluation} for an experimental setup and methods). Our profiling results reveal three major new insights.

\noindent\textbf{Insight 1:} \textit{Hot data that is used during application relaunch is usually similar between consecutive relaunches.}

\begin{figure}[!h]
\vspace{0.3em}
\centering
\includegraphics[width=0.48\textwidth]{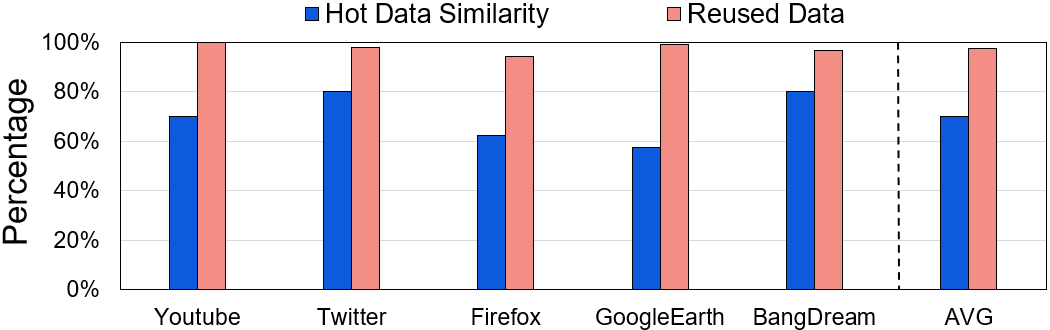}
\caption{\emph{Hot Data Similarity} and \emph{Reused Data} between two consecutive relaunches of an application across different applications.}
\label{fig:similarity}
\end{figure}

As discussed in Section~\ref{sec:Intro}, we categorize \emph{anonymous data} into three levels of hotness. Separating hot and cold data and treating them differently for compression and decompression can reduce relaunch latency and CPU usage by minimizing unnecessary compression, decompression, and swapping. 
To better separate hot and cold data, our profiler collects all data during an application's relaunches.  Each application is relaunched five times, and we collect hot, warm, and cold data for each relaunch. 
Figure~\ref{fig:similarity} shows the percentage of identical hot data between two consecutive relaunches of an application (i.e., \emph{Hot Data Similarity}) and the fraction of hot data from the prior relaunch that is reused in the later relaunch (i.e., \emph{Reused Data}). Hot Data Similarity is calculated by dividing the amount of identical hot data between two relaunches by the total hot data used during the second relaunch. Reused Data represents the percentage of hot data from the first relaunch that is present in the hot and warm data sets of the second relaunch. 

Based on the evaluation results, we make two observations. First, the average \emph{Hot Data Similarity} between two consecutive relaunches of an application is 70\%, indicating that hot data is generally similar between two consecutive relaunches of an application. Large \emph{Hot Data Similarity} exists because consecutive relaunches of an application typically involve starting the same activities and loading the same interface (e.g., application's logo, user interface) and other status information (e.g., game's status and user's status). 
Second, the average \emph{Reused Data} is 98\%, indicating that the hot data from one relaunch is highly likely to become the hot or warm data in the subsequent relaunch.

Hence, \textbf{the first key idea} of our design is to identify hot data using only the information from the most recent relaunch and manage data based on its identified hotness to reduce unnecessary compression and decompression. There are two challenges to realizing this first key idea: 1) How can we identify data hotness dynamically with low overhead (e.g., CPU time and energy consumption)? and 2) How can we effectively handle data based on varying levels of hotness, specifically by either keeping it uncompressed in main memory, compressing it in DRAM, or swapping it into flash memory-based swap space?
To effectively address these two challenges, we exploit two other new insights (Insights 2 and 3) that we explain next.

\noindent\textbf{Insight 2:} \textit{When compressing the same amount of data, a small-size compression approach (i.e., compressing data in small chunks) is much faster than a large-size compression approach at the cost of lowering compression ratio.}

Compression algorithms typically divide the entire application data into multiple chunks and compress each chunk separately~\cite{LZ4,LZO, mao2022trace}. These chunks can be multiple kilobytes or larger (i.e., \emph{large-size compression}) or multiple bytes to a few kilobytes of data (i.e., \emph{small-size compression}).

When compressing the same amount of data, large-size compression achieves a higher compression ratio than small-size compression, as it leverages redundancy over a broader data range~\cite{mahoney2011large}.  However, it is \emph{not} obvious whether large-size compression causes longer or shorter execution time as there are multiple conflicting factors affecting compression algorithm latency. For example, large-size compression better utilizes the memory bandwidth (due to loading large chunks of data) but at the same time has a larger memory footprint that can negatively affect cache performance~\cite{young2018cram,carvalho2021understanding}. To understand this trade-off better, we measure compression ratio and latency with varying chunk sizes (from 128B to 128KB) using mobile applications' \emph{anonymous data}. We use the default compression algorithms in Android systems, LZO~\cite{LZO} and LZ4~\cite{LZ4}, to compress 576 MB of \emph{anonymous data} from real applications included in Section~\ref{sec:back}.

\begin{figure}[!h]
\centering
\includegraphics[width=0.483\textwidth]{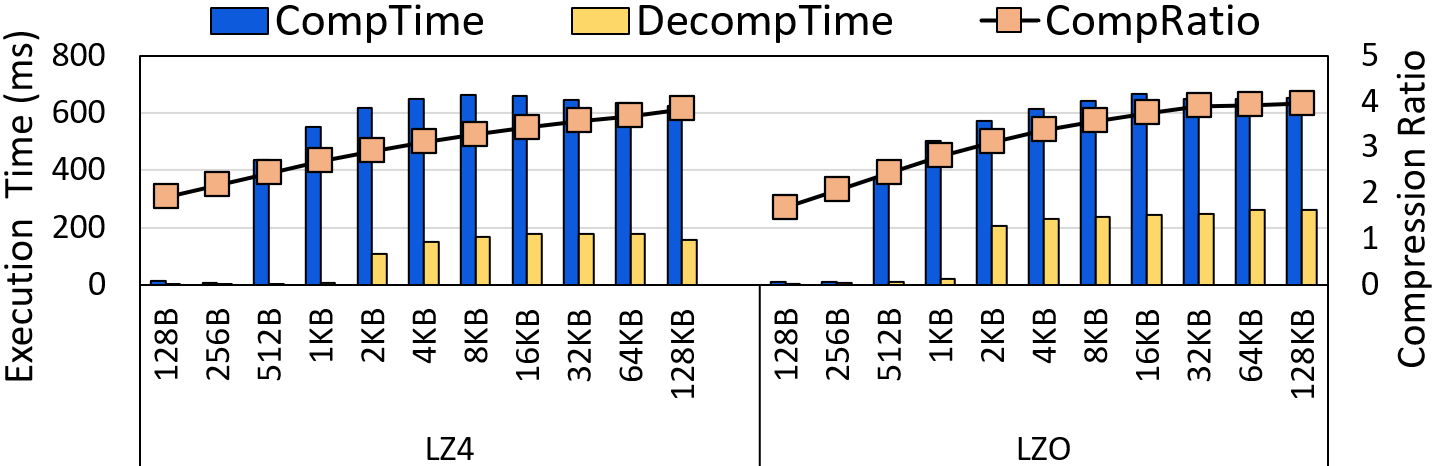}
\caption{Compression latency, decompression latency, and compression ratio under various compression chunk sizes. X-axis represents the compression chunk size. "128B" means 128 bytes of data is compressed per operation (i.e., a 4KB page will be compressed via 32 operations).}
\label{fig:comprresults}
\end{figure}

Figure~\ref{fig:comprresults} shows the compression latency (\texttt{CompTime}), decompression latency (\texttt{DecompTime}), and compression ratio (\texttt{CompRatio}) with various compression chunk sizes. Compression and decompression latencies indicate the time taken to compress and decompress a total of 576 MB. Compression ratio refers to the ratio of the original data size to the compressed data size and thus quantifies how much the data size is reduced using the compression algorithm.

We make two key observations. First, compression ratio increases from 1.7 to 3.9 as compression chunk size increases from 128B to 128KB.  Second, small-size compression is significantly faster for the evaluated mobile \emph{anonymous data} workloads.  For example, compression latency using 128B compression chunk size is 59.2$\times$ and 41.8$\times$ faster compared to using 128KB compression chunk size for LZ4 and LZO compression algorithms, respectively. The primary reason for faster small-size compression is the finer data granularity in our evaluated mobile workloads, such as Twitter, YouTube, and Firefox (see Section~\ref{sec:evaluation}). An \emph{anonymous page} contains multiple types of data blocks, and similar types of data are gathered within a small region (e.g., 128B or 512B), which increases the efficiency of small-size compression.\footnote{To foster further research in the design and optimization of mobile compressed swap techniques, we open-source our implementations at https://github.com/CMU-SAFARI/Ariadne.} 

Hence, \textbf{the second key idea} of our design is to employ different compression chunk sizes depending on the hotness level of data. For example, we compress cold data using larger compression chunk sizes to achieve a better compression ratio without worrying about slow decompression latencies, as cold data is unlikely to be read again. For hot data, on the other hand, we use small compression chunk sizes to reduce decompression latency.  
The challenge in effectively realizing this idea lies in correctly identifying the hotness level of data. Inaccurate identification of the data hotness level can impose both latency and memory capacity overheads. To mitigate the penalty of inaccurate identification, we introduce a new insight (Insight 3), which we explain next.

\noindent\textbf{Insight 3:} \textit{There is locality in the address space (i.e., sector numbers) in \emph{zpool} when swapping anonymous data into main memory during application relaunch.} 

To mitigate the penalty of inaccurately identifying data hotness levels, we aim to hide decompression latency and the latency of swapping data into main memory (called swap-in) by decompressing and swapping soon-to-be-used data in advance. To do this effectively, we need to predict the next set of data to be used. We assess the spatial locality in accesses to the compressed pages in \emph{zpool}. To this end, we measure the probability of accessing \emph{N} \emph{consecutive pages} (i.e., pages that are physically adjacent in \emph{zpool}). 
Table~\ref{tab:locality} reports the probability of accessing \emph{two} or \emph{four} \emph{consecutive pages} in \emph{zpool} for each evaluated application.

\begin{table}[h!]
\centering
\caption{The probability
of accessing two or four \emph{consecutive pages} in \emph{zpool} for each
evaluated application.}
\footnotesize
\begin{tabular}{c|c|c|c|c|c}
\hline
 &\textbf{Youtube}& \textbf{ Twitter}&\textbf{Firefox}&\textbf{GoogleEarth}&\textbf{BangDream}\\
 \hline
 \hline
 2&0.86&0.81&0.69&0.77&0.61\\
 \hline
 4&0.72&0.61&0.43&0.54&0.33\\
\hline
\end{tabular}
\label{tab:locality}
\end{table}

We make two key observations. First, most applications exhibit locality in data access in  \emph{zpool} when swapping in \emph{anonymous data} during application relaunch. For example, the probability of accessing two \emph{consecutive pages} is 86\% for YouTube. This means that if we pre-decompress and pre-swap the immediate next page of the currently-being-accessed page, 
the pre-swapped page has an 86\% chance of being used by the application soon. 
Second, the probability of accessing four \emph{consecutive pages} is significantly lower than that of two \emph{consecutive pages} (17\%-46\% lower across various applications). Hence, pre-decompressing the three immediate next pages can pollute main memory with pages that are \emph{not} going to be used.

Hence, \textbf{the third key idea} of our design is to predict the next set of data to be used and pre-decompress it, reducing the impact of swapping data back into main memory and decompression latency on application relaunch. There are two design decisions associated with realizing this idea: 1) How much data should be pre-decompressed? and 2) When should we do pre-decompression? Our design addresses these decisions, as presented in Section~\ref{sec:mechanisim}.

\noindent\textbf{Summary.} We uncover three new insights by analyzing modern mobile workloads. First, hot data is usually similar between consecutive application relaunches.
Second, small-size compression/decompression is fast, while large-size compression achieves a better compression ratio.
Third, there is locality in data access in  \emph{zpool} when swapping in \emph{anonymous data} during application relaunch.
These new insights lead to three key ideas: hotness-awareness data organization, size-adaptive compression, and pre-decompression, as we discuss in Section~\ref{sec:mechanisim}. 

\section{\proposal Design}
\label{sec:mechanisim}

\subsection{Design Overview}
We propose \proposal, a new compressed swap scheme for mobile devices that reduces application relaunch latency and CPU usage while increasing the number of live applications for enhanced user experience.
The key idea of \proposal is to reduce the frequency and latency of compression, decompression, swap-in, and swap-out operations by leveraging different compression chunk sizes based on the hotness level of the data, while also performing speculative decompression based on data locality characteristics.

\begin{figure}[!h]
\centering
\vspace{-3pt}
\includegraphics[width=0.96\linewidth]{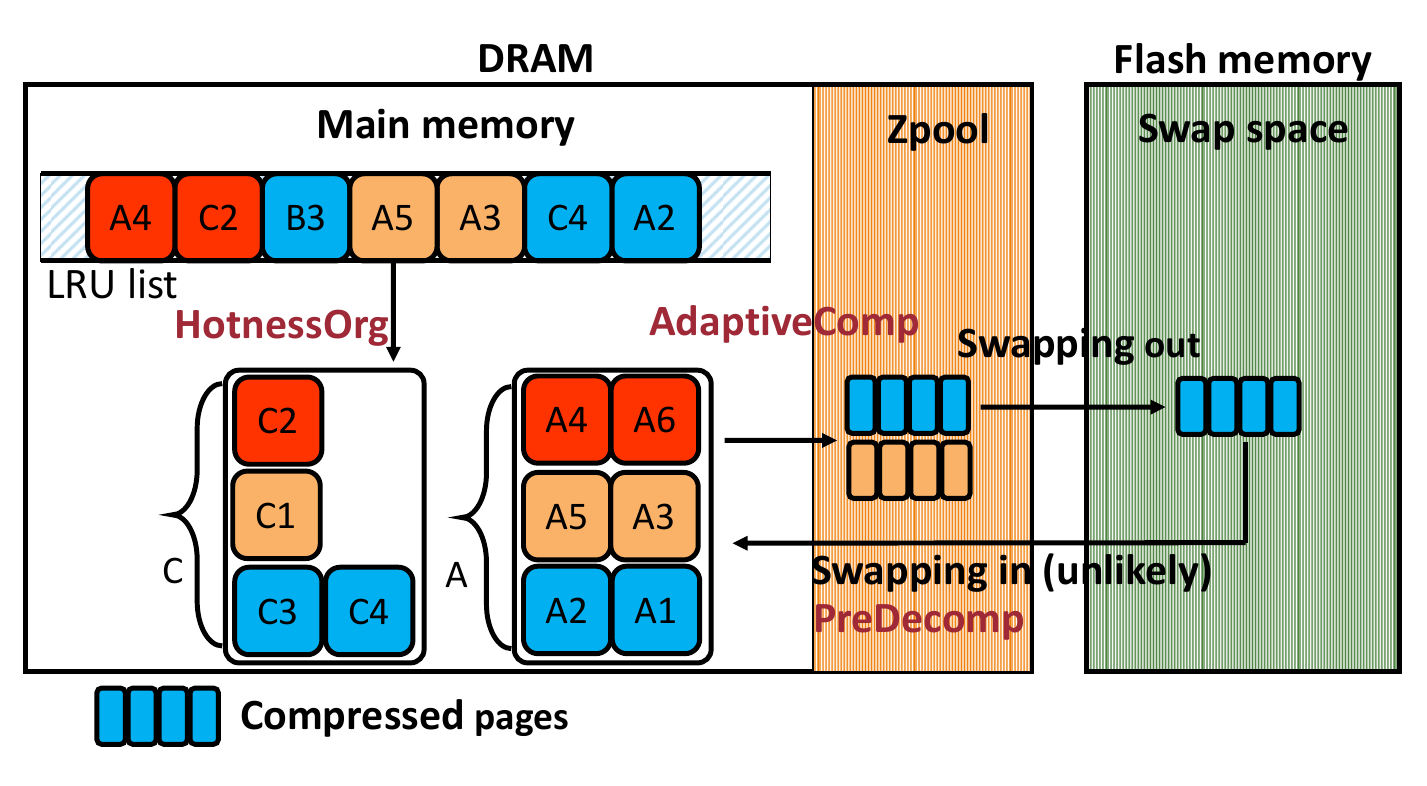}
\caption{Design overview of \proposal. \proposal incorporates three key techniques: \dataorg, \compress, and \predi. It involves  \emph{zpool} and flash memory-based swap space management. Blocks refer to data pages in main memory. Colors represent the hotness of the data
pages: red (hot), orange (warm), blue (cold).}
\label{fig:overview}
\end{figure}

\noindent\textbf{Data storage architecture of \proposal.} Figure~\ref{fig:overview} presents an overview of \proposal. We use colors to represent the hotness levels of data pages: red for hot, orange for warm, and blue for cold data. In Android systems, \emph{anonymous data} of running applications can be stored in main memory,  \emph{zpool}, or flash memory-based swap space. Main memory has the lowest access latency, while flash memory-based swap space has the highest. Therefore, systems usually prioritize storing \emph{anonymous data} in main memory for best performance. When main memory capacity is limited, systems use the  \emph{ZRAM} scheme to compress the least recently used (LRU~\cite{LRU}) data into  \emph{zpool}. 
The flash memory-based swap space serves as main memory extension to store compressed data swapped out from  \emph{zpool} when there is insufficient main memory space. \proposal chooses to swap out \emph{compressed} data,  which leads to smaller writes to flash memory and lower storage space consumption.  However, this design choice may increase read latency due to decompression. 
We reduce the probability of incurring such latency by mainly writing cold data (that is unlikely to be read again) into the flash swap space. 

\noindent\textbf{Key mechanisms of \proposal.} 
Based on the above data storage architecture, \proposal incorporates three techniques: 
First, \proposal uses a low-overhead, hotness-aware data organization mechanism, called \dataorg, to determine data hotness and maintain data with different levels of hotness in separate memory page lists accordingly. 
The goal of \dataorg is to reduce the frequency of compression/decompression and swap-in/swap-out operations. To achieve this goal, \proposal aims to maintain uncompressed hot data in main memory,  compress warm data into \emph{zpool}, and swap compressed cold data to the flash memory-based swap space (see Section~\ref{sec:dataorg}).
Second, \proposal enables a size-adaptive compression mechanism, called \compress, to leverage the benefits of different compression chunk sizes. The goal of  \compress is to achieve both short relaunch latency and a good compression ratio by using small-size compression chunks for identified warm data and large-size compression chunks for cold data (see Section~\ref{sec:compress}). 
Third, rather than relying on on-demand decompression or data swapping-in operations during application relaunches, \proposal employs a \emph{proactive and predictive decompression} (i.e., \emph{predecompression}) mechanism, called \predi, that leverages data locality to proactively determine the best data and timing for compression and swapping. The goal of \predi is to mitigate the negative impact of read latency on the user experience (see Section~\ref{sec:ecivtion}). 
We also consider the compatibility of \proposal with different compression algorithms and memory management optimizations (See Section~\ref{sec:implement}).

\subsection{Low-overhead Hotness-Aware Data Organization}
\label{sec:dataorg}

\begin{figure}[!h]
\centering
\includegraphics[width=0.487\textwidth]{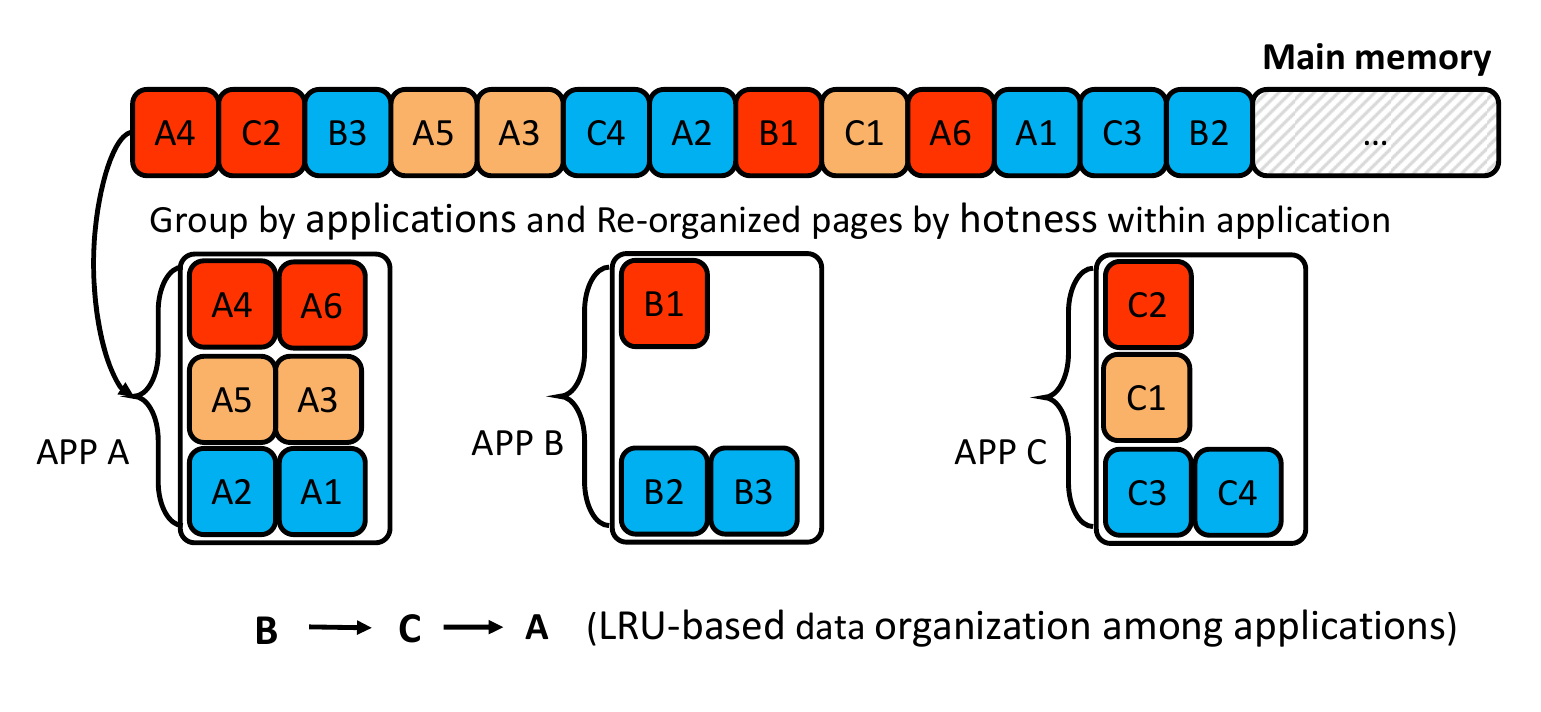}
\caption{Hotness-aware data organization (\dataorg). Blocks refer to data pages in main memory. Colors represent the hotness of the data
pages:  red (hot), orange (warm), blue (cold).}
\label{fig:dataorg}
\end{figure}

We propose a hotness-aware data organization mechanism, called \dataorg, that builds on LRU-based memory management. 
\dataorg aims to improve compression efficiency by separating hot, warm, and cold data efficiently. The challenge is how to identify data hotness dynamically and accurately with low memory capacity and CPU overhead.
Specifically, \dataorg encompasses two aspects: data organization within an application and data organization among applications, as shown in Figure~\ref{fig:dataorg}.

\noindent\textbf{Within an application.}  Data organization within an application involves three components: hotness initialization, hotness update, and data eviction. \dataorg separates all the anonymous data of the application into three LRU lists (hot, warm, and cold) rather than the typical two lists (active and inactive by default~\cite{Linuxkernel, LRUactive}).
First, for hotness initialization,  when a system launches an application for the first time, the system adds a certain amount of data used during the launch to the hot list (i.e., the LRU list to store hot data in main memory). To reduce overhead,  we profile data usage for each application during its relaunch to determine the \emph{initial} size (i.e., data amount) of the hot list. The profiling procedure for this size is the same as the one used for the data shown in Figure~\ref{fig:similarity}. This profiling works effectively because: i) the amount of hot data remains similar for each relaunch of an application, as shown in our collected traces and the results in prior work~\cite{son2021asap}; ii) \proposal adaptively updates the hot list during application relaunch and execution. 
Then, the system adds other data generated during application execution to the cold list. If the application accesses data in the cold list during execution, the system moves the data to the warm list. Moving data from the cold list to the warm list is similar to default Android systems, which move data from the inactive list to the active list.  This initialization procedure does not incur additional overhead.

Second, for hotness update (i.e., moving data among hot, warm, and cold lists according to the access pattern), after relaunching an application, the system moves all old data in the hot list to the warm list and adds the data from this relaunch to the hot list. This ensures that the hot data from the most recent relaunch is in the hot list.
Third, for data eviction, the system first chooses data from the cold list of an application for compression. If all cold data of all applications are compressed, it starts compressing data from the warm list, and finally (if absolutely necessary) the hot data. When the  \emph{zpool} space is insufficient to store all the compressed data, the system writes some compressed data to flash memory-based swap space following a policy that ensures cold data is swapped out first.

\noindent\textbf{Across applications.} For data organization across applications, we have two policies. First, we continue using the LRU policy to manage an LRU-based application list. Applications are added to the LRU list based on the access time of their most recently accessed page (eviction order of applications is A, C, and B, as shown in Figure~\ref{fig:dataorg}). Second, we prioritize using the main memory (DRAM) capacity for foreground applications. This policy is compatible with the $mem\_cgroup$ function~\cite{memcgroup} that can be enabled in the Linux kernel.

All data organization tasks involve only LRU list operations, without physically moving data, similar to the baseline system. Only adding old hot data to the warm list is an additional LRU operation compared to the baseline system. Thus, \dataorg is a low-overhead data organization mechanism.

In summary, \dataorg efficiently identifies and exploits data hotness.
By leveraging data access patterns during application relaunch and execution, \dataorg can efficiently identify data hotness and manage it with minimal overhead.

\subsection{Efficient Size-Adaptive Compression}
\label{sec:compress}

We propose an efficient size-adaptive memory compression mechanism, called \compress, that allows for compressing data using different compression chunk sizes based on data hotness.

\begin{figure}[!h]
\centering
\includegraphics[width=0.483\textwidth]{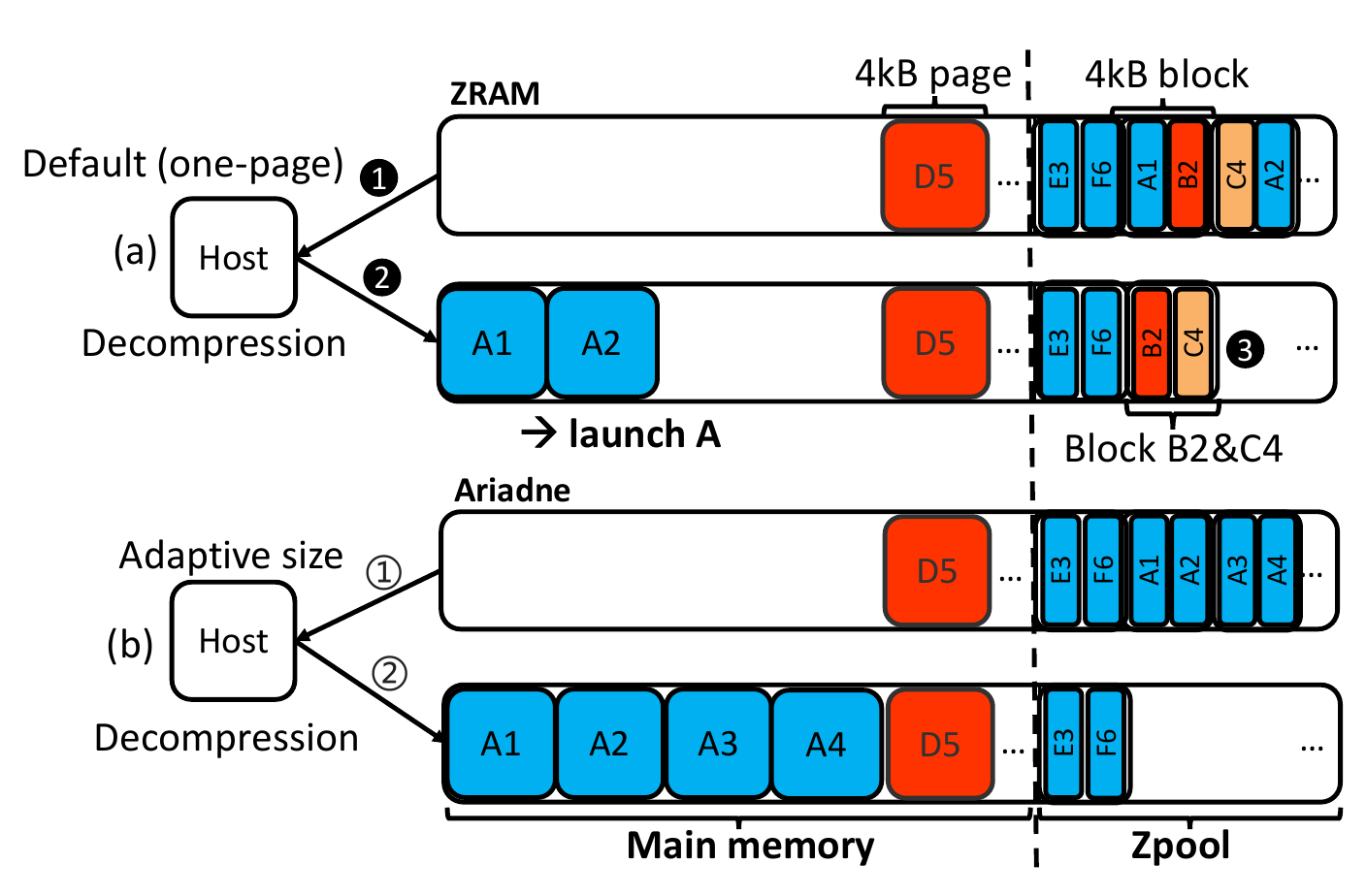}
\caption{Decompression procedure of efficient size-adaptive compression (\compress). Colors represent the hotness of the data pages: red (hot), orange (warm), blue (cold). The data layout in \emph{zpool} differs between the baseline \emph{ZRAM} scheme and \proposal, as they employ different data organization policies.}
\label{fig:large-size}
\end{figure}

\noindent\textbf{Adaptive size according to data hotness for data compression}. Large-size chunks for compression (i.e., large-size compression) are not commonly used in current mobile systems for two reasons. 
First, large-size compression tends to increase data movement, computational overhead, and energy consumption, as large data chunks could involve more unused data, which could be redundantly transferred between host CPU and DRAM, as shown in Figure~\ref{fig:datamove}. 
Second, according to our Insight 2 in Section~\ref{sec:insights}, the compression and decompression latency is longer when using large-size compression.
\compress addresses these issues by leveraging the hot and cold data separation supported by \dataorg (see Section~\ref{sec:dataorg}). 
\proposal utilizes large-size compression for \emph{cold} data to achieve a good compression ratio.
Compressing only cold data using large chunks mitigates performance penalties of using large chunks since cold data is unlikely to be reused. 
Conversely, \compress uses small-size compression for hot and warm data to achieve better relaunch latency and execution performance. As a result, \compress can take advantage of different compression chunk sizes without incurring their typical penalties. 
Figure~\ref{fig:large-size} compares the decompression procedures for a given compressed page in \emph{ZRAM} versus \compress. We use decompression as an example to explain \compress's workflow, as it better illustrates the penalties and benefits of our design compared to the baseline compression mechanism used in \emph{ZRAM}.

For the baseline one-page (i.e., 4KB) compression chunk size (see Figure~\ref{fig:large-size} (a)), when a user launches application A, the system reads the A-related compressed blocks (blocks A1\&B2 and C4\&A2) from  \emph{zpool} to the host CPU \ding{182} and decompresses them. The system writes the decompressed pages A1 and A2 back to main memory (DRAM) to facilitate application A's launch \ding{183}. Finally, the system merges the unused compressed data and writes block B2\&C4 back to  \emph{zpool} \ding{184}.
When decompressing block A1\&B2, B2 is \emph{not} decompressed because the system uses a one-page compression chunk size, meaning pages A1 and B2 are compressed individually.

In contrast, the decompression procedure in \proposal differs from that in the default \emph{ZRAM} in two major ways:
First, \proposal's \dataorg organizes data based on its hotness level, unlike the default \emph{ZRAM} that uses LRU policy. As a result, the data layout in both the main memory and the \emph{zpool} using \proposal differs significantly from that in \emph{ZRAM}. For example, when a user launches application A, the hot data required for the relaunch is in main memory. 

Second, \proposal performs compression operations based on data hotness levels. For example, large-size compression targets cold data that is unlikely to be accessed again. To illustrate both the benefits and potential drawbacks of large-size compression, we present a worst-case scenario (i.e., need to decompress the data that was compressed using a large size) of large-size decompression using \proposal in Figure~\ref{fig:large-size} (b). The worst case occurs when cold data is incorrectly predicted and later needs to be used after being compressed. In this case, when the system reads the A-related compressed blocks (A1\&A2 and A3\&A4) from  \emph{zpool} to the host CPU and decompresses them \ding{172}. By leveraging a large-size compression policy, the system decompresses pages A1 and A2 together, as well as A3 and A4 together. Following decompression, the system writes all four decompressed pages back to the main memory space in DRAM \ding{173} to facilitate application A's relaunch. 
Thus, when compressing data at large granularity, the system decompresses compressed blocks (e.g., A1\&A2 and A3\&A4) entirely, even if the application only requires a small portion of the blocks. This can result in wasted CPU time and memory capacity if decompressed pages are not accessed together.

By leveraging the hotness level-based data separation provided by \dataorg, \proposal enables an efficient size-adaptive compression mechanism, \compress. \compress enables the processing of multiple cold data pages using a single compression operation while avoiding the drawbacks of compressing a large amount of data at once. Consequently, \proposal significantly reduces the frequency of data compression and decompression operations by leveraging \dataorg and \compress, thereby lowering application relaunch latency and CPU usage. To further mitigate the impact of decompression latency on application relaunches, we propose hiding this latency by decompressing soon-to-be-used data in advance, which we explain next.

\subsection{Proactive and Predictive Decompression}
\label{sec:ecivtion}
We propose a proactive decompression mechanism, called \predi  
(i.e., pre-decompression), to efficiently and proactively perform decompression operations ahead of reading, thereby reducing the negative impact of decompression latency on read latency. The challenge lies in accurately predicting the best data and timing for decompression while minimizing CPU and memory capacity overhead.

\noindent\textbf{Fast prediction of data to be decompressed.}
According to Insight 3 in Section~\ref{sec:insights}, there is locality in data access in  \emph{zpool} when swapping-in \emph{anonymous data} during application relaunch. \dataorg organizes and maintains all hot and warm data with high locality. Thus, \predi can accurately predict the next data for decompression using the data layout organized by \dataorg. For example, when a compressed page is required, its subsequent page will also be proactively pre-decompressed. Since the probability of accessing two \emph{consecutive pages} is high (see Table~\ref{tab:locality}), we pre-decompress only one compressed page at a time, ensuring high accuracy while minimizing the memory capacity overhead required to store the pre-decompressed data.

\noindent\textbf{Lightweight pre-decompression method.} To support pre-decompression, \proposal maintains a buffer in the main memory to store the pre-decompressed data. When the buffer is full, it uses a first-in, first-out policy. The larger the buffer size, the longer compressed data can be stored before it is used in main memory. For example, if the buffer size is only one page, the system should use the pre-decompressed data immediately. Otherwise, the data will be compressed again. 

In summary, \predi exploits data access patterns to determine the best time and data for which to perform decompression and swapping. By doing so, \predi efficiently performs pre-decompression operations to mitigate read latency, thereby enhancing user experience.

\subsection{Compatibility of \proposal with Other Techniques}
\label{sec:implement}
We discuss the compatibility of \proposal with: i) different compression algorithms and ii) other memory management schemes (e.g., memory allocation and memory reclamation schemes).

\noindent\textbf{Compression algorithms.}
It is valuable for systems to support multiple compression algorithms to cater to different objectives. Therefore, ensuring compatibility with a range of compression algorithms is important.
\proposal is compatible with various compression algorithms, such as  LZO~\cite{LZO}, LZ4~\cite{LZ4}, and base-delta compression{\cite{pekhimenko2012base, pekhimenko2013linearly}.
\proposal naturally supports different compression algorithms, such as switching between LZO and LZ4, as it inherits \emph{ZRAM}'s interface.  Compression algorithms might require slight interface modifications to support \compress by adjusting the compression chunk size.

\noindent\textbf{Impact on other memory management schemes.}
\proposal is compatible with various baseline memory management algorithms, such as default memory allocation and memory reclamation techniques used in modern systems. \proposal does \emph{not} impact the memory allocation procedure because memory allocation operates only on available memory space, while \proposal focuses on organizing data.  However, \proposal affects the memory reclamation scheme. Specifically, in a mobile system, the memory reclamation scheme selects data to reclaim (called victim data) based on the LRU policy. In contrast, the memory reclamation scheme of \proposal uses our hotness-aware data organization scheme (\dataorg) to select victim data, enhancing reclamation efficiency by prioritizing the reclamation of cold data.
\proposal is also compatible with swap scheme optimizations, such as MARS~\cite{guo2015mars}, FlashVM~\cite{saxena2010flashvm}, Fleet~\cite{end2024more}, and SmartSwap~\cite{zhu2017smartswap}, as \proposal is complementary to them.

\section{Evaluation Methodology}
\label{sec:evaluation}

\noindent\textbf{Experimental platform.} 
We implement and evaluate \proposal on a real commercial smartphone,
Google Pixel 7~\cite{Pixel7} with the latest Android 14 operating system~\cite{Android14}.
We list the detailed real system configuration in Table~\ref{tab:phones}.

\begin{table}[h!]
\centering
\caption{Experimental Platform Configuration.}
\footnotesize
\begin{tabular}{c|c|c}
\hline
\textbf{Name} & \textbf{System}& \textbf{Memory \& Storage} \\ \hline
\hline
 & CPU: 8 cores  & 12GB DRAM\\
Google  &2x 2.85 GHz Cortex-X1 ~\cite{dempsey2021reviews} \& &\\
Pixel 7 &2x 2.35 GHz Cortex-A78 ~\cite{al2023comparative} \& & 128GB flash\\
&4x 1.8 GHz Cortex-A55~\cite{seo2020optimized} & \\
& Android: 14; Linux 5.10.157~\cite{Android14} & UFS3.1\\
\hline
\end{tabular}
\label{tab:phones}
\end{table}

\noindent\textbf{Workloads.}
We execute ten 
popular applications (Twitter, YouTube, TikTok, Edge, Firefox, Google Earth, Google Maps, BangDream, Angry Birds, and TwitchTV) via MonkeyRunner~\cite{monkey} to collect mobile workload traces.\footnote{Recent studies~\cite{liang2020acclaim, liang2022cachesifter} show that mobile users often run more than eight applications concurrently.}
Using mobile workload traces makes our methodology and results reproducible, as opposed to running real applications that execute differently for each different test. We use the collected traces for both insight analysis and final evaluation results. For example, we use the collected page data in traces as the input of compression and decompression algorithms for both the state-of-the-art \emph{ZRAM} scheme and \proposal. This allows us to reproducibly and consistently compare their compression latency, decompression latency, and compression ratio.
A trace is composed of the page frame number (PFN),  \emph{ZRAM} sector, source application number (UID), and page data that needs to be compressed, swapped-in or swapped-out.
We create \emph{ten} traces. Our procedure for creating each trace is as follows. First, for each trace, we select a target application out of ten applications to launch and execute. Second, we put the target application in the background and launch the other nine applications. To capture more information across various usage scenarios, we launch the nine applications in different orders, creating several (e.g., three) distinct usage scenarios for each target application. Third, we relaunch the target application to collect its relaunch information.

To prevent interference during each trace creation and ensure the reproducibility of our methodology, we perform the following three actions for each trace collection: i) we close all applications and clear their cache files before rebooting the smartphone to eliminate the impact of old cache files, ii) after rebooting the smartphone, we clean the cache again to eliminate the impact of potentially buffered data in main memory, and iii) we use the same applications with the same user account and perform the same sequence of activities using an auto-testing script via MonkeyRunner~\cite{monkey} to avoid human bias.

To foster further research in the design and optimization of mobile compressed swap techniques, we open-source all our source code, traces, and scripts at https://github.com/CMU-SAFARI/Ariadne.

\noindent\textbf{Evaluated Schemes.} We evaluate two compressed swap schemes:
\noindent\textbf{ 1)} \emph{ZRAM}~\cite{zram1, zram2, new-zram, merge-zram}, which is the state-of-the-art compressed swap scheme used in modern Android systems. ZRAM employs Least Recently Used (LRU)~\cite{LRU} as the default policy for selecting data to compress. With LRU, the system selects the least recently used pages for compression. Modern Android systems optimize memory page organization by grouping data based on the associated application. This solution only supports single-page-size (i.e., 4KB) compression to avoid potential penalties (as discussed in Section~\ref{sec:compress}) and does  \emph{not} allow data to be decompressed \emph{before} it is required by the system, avoiding memory capacity waste and unnecessary CPU usage if the decompressed data will not be used.

\noindent\textbf{2)} Different versions of \proposal, which is our proposed compressed swap scheme. We evaluate \proposal under different configurations, whose parameters are shown in Table~\ref{table:3}.
$S$ represents the size of  \emph{zpool}, which is set to 3GB. This parameter determines the maximum number of compressed pages that can be stored in  \emph{zpool} and consequently affects user experience in two ways: 1) The size of \emph{zpool} impacts the number of writes to the NAND flash memory 
and its overall lifetime, and 2) it affects the relaunch-related data placement, thereby impacting application relaunch latency.
$Small Size$, $Medium Size$, and $Large Size$ represent the compression chunk sizes for hot list, warm list, and cold list, respectively. They serve as inputs for compression algorithms and affect both the compression ratio and compression latency. We denote these size configurations as $Small Size$\--$Medium Size$\--$Large Size$ (e.g., 1K-2K-16K) for each version of \proposal in Section~\ref{sec:evel}.

\begin{table}[h!]
\vspace{0.5em}
\centering
\caption{Summary of parameters used by \proposal.}
\label{table:3}
\footnotesize
\begin{tabular}{c|c|c}
  \hline
  \textbf{Parameter}          & \textbf{ Description} & \textbf{Setting (B)} \\  \hline \hline
  $S$ & Size of ZRAM partition& 3G \\  \hline 
  $Small Size$ & Compression chunk size for hot list & 256,512,1K \\  \hline
  $Medium Size$ & Compression chunk size for warm list & 2K,4K \\  \hline
  $Large Size$ & Compression chunk size for cold list & 16K,32K \\  \hline
\end{tabular}
\end{table}

During an application relaunch, the system fetches all the launch-related data into main memory. In the optimal case, all data resides in main memory (DRAM). In other scenarios, we consider two situations: i) where data in the \emph{hot} list is in main memory while other data is in either \emph{ZRAM} or flash memory-based swap space, and ii) where all data needs to be read from either \emph{ZRAM} or flash memory-based swap space. The first scenario excludes hot list data from compression and decompression operations, and the second scenario applies compression and decompression on data in all lists. We abbreviate these two scenarios as EHL (exclude hot list) and AL (all lists).

\noindent\textbf{Evaluated Metrics.}
We evaluate \proposal using three major sets of metrics.
First, we assess the impact of \proposal on factors that influence user experience, including application relaunch latency and the CPU usage due to compression and decompression.
Second, we analyze the auxiliary metrics, including compression/decompression latency, compression ratio, and the accuracy and coverage of data hotness level identification.
Third, we analyze memory capacity and CPU usage associated with our full \proposal implementation (not just compression/decompression).
\section{Evaluation Results}
\label{sec:evel}
We evaluate the effectiveness of \proposal compared to the state-of-the-art \emph{ZRAM}. Section 6.1 shows the overall effect of \proposal on the user experience with modern mobile devices. Section 6.2 analyzes the effectiveness of key techniques of \proposal by presenting auxiliary metrics.
Section 6.3 provides a sensitivity study on compression chunk size configurations. Section 6.4 studies the memory capacity and CPU usage associated
with our full \proposal implementation.

\subsection{Effect on User Experience} 
\label{sec:userexperience}
 
There are two metrics that significantly affect the user experience on mobile devices: i) application relaunch latency~\cite{end2024more,lebeck2020end} and ii) CPU usage, which directly impacts battery usage~\cite{pramanik2019power}.

\noindent\textbf{Application relaunch latency.} 
Figure~\ref{fig:hot-launch-latency} shows the application relaunch latency of the evaluated applications under different compressed swap schemes (i.e., \emph{ZRAM} and \proposal with different configurations).  We implement Ariadne on a real smartphone, the Google Pixel 7~\cite{Pixel7}, running the Android 14 operating system~\cite{Android14}. We execute our traces that are collected under different data organization policies (i.e., LRU in baseline \emph{ZRAM} and \dataorg in \proposal) on the smartphone using the \emph{ZRAM} scheme and \proposal, respectively. To demonstrate a lower bound for the best possible (optimal) latency for application relaunches, we also evaluate application relaunch latency under an ideal scenario, called \emph{DRAM}, where the system reads \emph{all} application data directly from DRAM (with the optimistic assumption that DRAM is large enough to host all such data), i.e., there is no swapping overhead. The x-axis represents the evaluated applications, and the y-axis indicates their relaunch latencies.
We report results for five randomly selected applications (out of 10) for readability.\footnote{ We present the same applications for Figures~\ref{fig:hot-launch-latency}–\ref{fig:compression-ratio-result} and Figure~\ref{fig:sensitivity}. We release the results for all applications in our GitHub repository~\cite{github} and the appendices of the extended version of the paper~\cite{LiangAxiv}.}

\begin{figure}[!h]
\vspace{0.3em}
\centering
\includegraphics[width=0.478\textwidth]{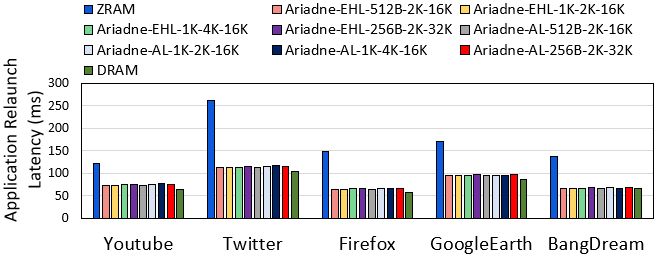}
\caption{Application relaunch latency.}
\label{fig:hot-launch-latency}
\end{figure}

We make two key observations. First, all versions (i.e., different configurations) of \proposal reduce the relaunch latency by around 50\%, on average, compared to \emph{ZRAM}. We believe \proposal enhances the user experience by reducing the relaunch latency as users switch among various applications with high frequency (e.g., >100 times a day~\cite{deng2019measuring}).
 Second, the relaunch latency of all Ariadne configurations is within 10\% percent of that of the optimistic DRAM configuration. This demonstrates that Ariadne effectively hides most of the latency due to compressed swapping in main memory.
Third, the performance difference between EHL and AL is negligible for a given same-size configuration. For instance, YouTube's relaunch latencies are 73 ms and 75 ms under \emph{Ariadne-AL-1K-2K-16K} and \emph{Ariadne-EHL-1K-2K-16K}, respectively. This is because \proposal intelligently adapts to different compression chunk sizes based on the hotness level of the data.
We conclude that \proposal effectively reduces application relaunch latency, thereby significantly enhancing user experience.

\noindent\textbf{CPU usage.}
Figure~\ref{fig:compression-power-result} illustrates the CPU usage for compression and decompression procedures across different versions of \proposal, normalized to the CPU usage for these procedures using the baseline \emph{ZRAM} scheme.

We make three key observations. First, all versions of \proposal with EHL significantly reduce CPU usage during compression and decompression for applications that generate more hot data. For instance, \proposal with EHL reduces CPU usage by 25\% for YouTube and 30\% for Twitter. 
Second, \proposal with AL, using smaller-size compression (e.g., \emph{256B-2K-32K}) achieves similar CPU usage to \proposal with EHL. Third, for the applications that produce less hot data\footnote{ We report the proportion of data at different hotness levels from traces in our GitHub repository~\cite{github} and the appendices of the extended version of the paper~\cite{LiangAxiv}.}, such as BangDream, CPU usage increases by about 3\% with EHL versus with AL. This is because more data is compressed using larger sizes.
However, since warm and cold data are accessed less frequently than hot data, \proposal can effectively offset the CPU overhead caused by large-size compression by reducing the frequency of compression and decompression operations.

Overall, compared to the state-of-the-art \emph{ZRAM}, \proposal achieves an average CPU usage reduction of approximately 15\% across all configurations.  We conclude that \proposal significantly reduces CPU usage compared to the baseline \emph{ZRAM} scheme that underscores the effectiveness of \dataorg and \compress.

\begin{figure}[!h]
\vspace{0.3em}
\centering
\includegraphics[width=0.475\textwidth]{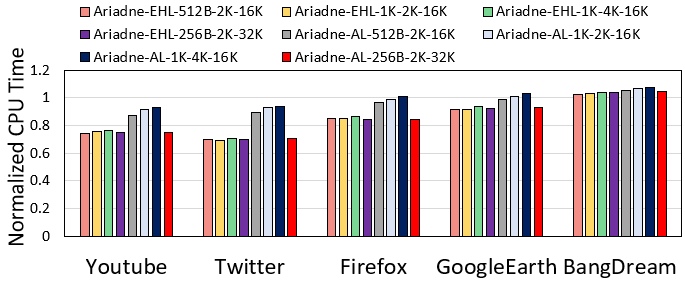}
\caption{Normalized CPU usage of compression and decompression procedures across different versions of \proposal, normalized to the CPU usage for these procedures under \emph{ZRAM}.}
\label{fig:compression-power-result}
\end{figure}

\subsection{Analysis of \proposal}
\label{sec:breakdown}
To study the effectiveness of the key techniques of \proposal, we investigate how \proposal influences relaunch latency and CPU usage through four key auxiliary metrics: i) compression and decompression latency, ii) compression ratio, iii) accuracy and coverage of hot data identification for an application relaunch.

\noindent\textbf{Compression and decompression latency.}  Figure~\ref{fig:decompression-latency-result} shows the data compression and decompression latency in evaluated applications. We evaluate LZO~\cite{LZO} compression algorithm supported by Google Pixel 7. The x-axis represents the evaluated applications, and the y-axis shows the compression and decompression latency of data from their traces. 
 
\begin{figure}[!h]
\centering
\includegraphics[width=0.475\textwidth]{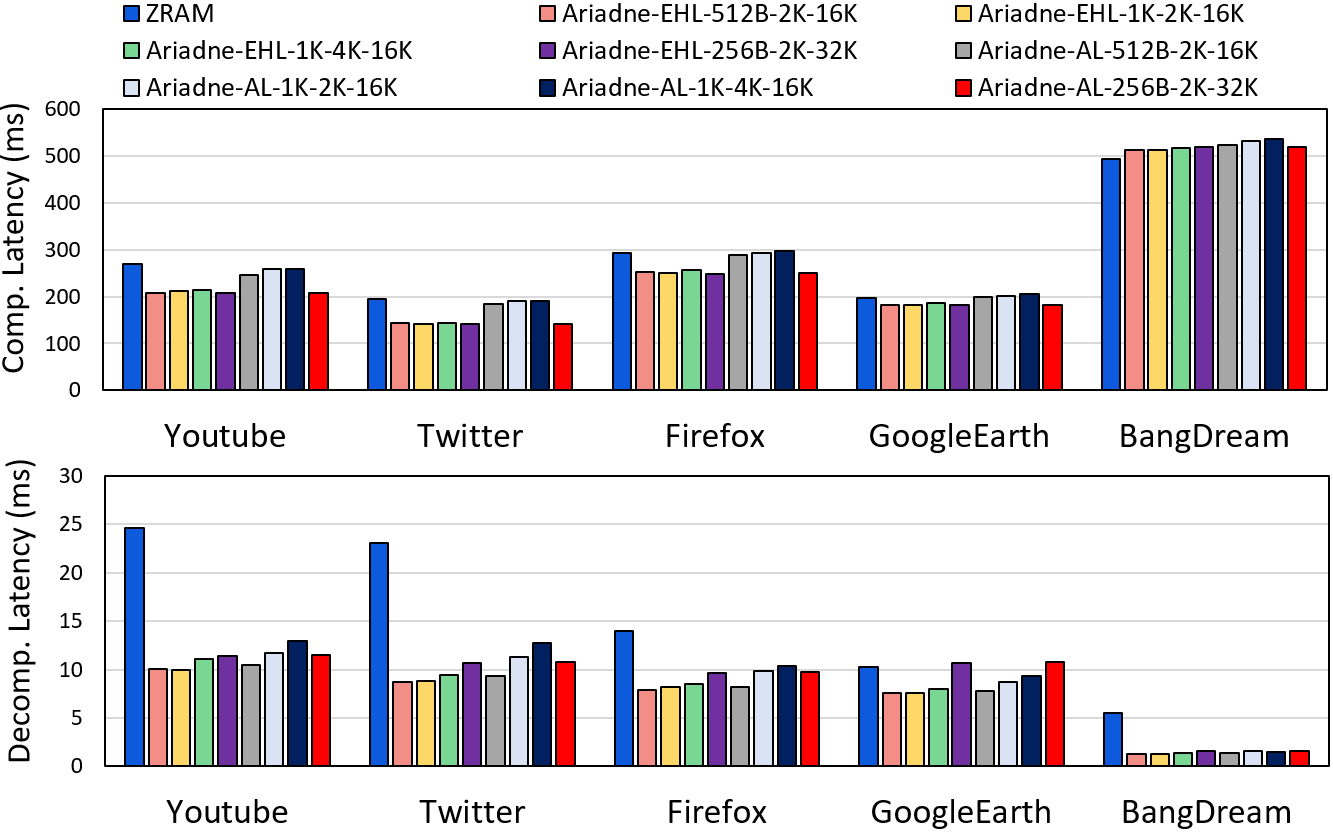}
\caption{Compression and decompression latency using different compressed swap schemes (i.e., different versions of \proposal and \emph{ZRAM}).}
\label{fig:decompression-latency-result}
\end{figure}

We make two key observations.
First, all versions of \proposal\ significantly reduce the decompression latency. For example, \proposal\ with the configuration \emph{1K-2K-16K} reduces the decompression latency by approximately 60\% for YouTube and Twitter, and by approximately 90\% for BangDream, compared to the baseline \emph{ZRAM} scheme. This is because  \proposal uses fast, small-size compression on frequently decompressed data, i.e.,  hot and warm data, which leads to fast decompression. 
Second, compression latency is also reduced for all applications except BangDream. For YouTube and Twitter, \proposal-EHL with the configuration \emph{1K-2K-16K} reduces compression latency by 20\%. This reduction is primarily due to the use of large-size compression on cold data and reduced compression operations on hot data.
We conclude that \proposal\ significantly reduces decompression latency across various applications, which in turn reduces application relaunch latency.

\noindent\textbf{Compression ratio.}  Figure~\ref{fig:compression-ratio-result} presents the compression ratio of data of different applications when using different compressed swap schemes. 
We make two observations. First, \proposal-EHL with the size configuration \emph{1K-4K-16K} consistently provides better compression ratio than ZRAM for every application. This is because larger compression chunk sizes result in better compression ratios across all hotness levels of data (as we have discussed in Section~\ref{sec:insights}). 
Second, \proposal-AL, using smaller compression chunk sizes (i.e., \emph{512B-2k-16K}) achieves a similar compression ratio to that of \emph{ZRAM}. This is because we select size configurations to balance the tradeoff between compression and decompression latency and the compression ratio. We conclude that \proposal provides comparable or even better compression ratios, compared to the baseline ZRAM scheme, which can positively affect both application relaunch latency and flash memory lifetime.

\begin{figure}[!h]
\centering
\includegraphics[width=0.47\textwidth]{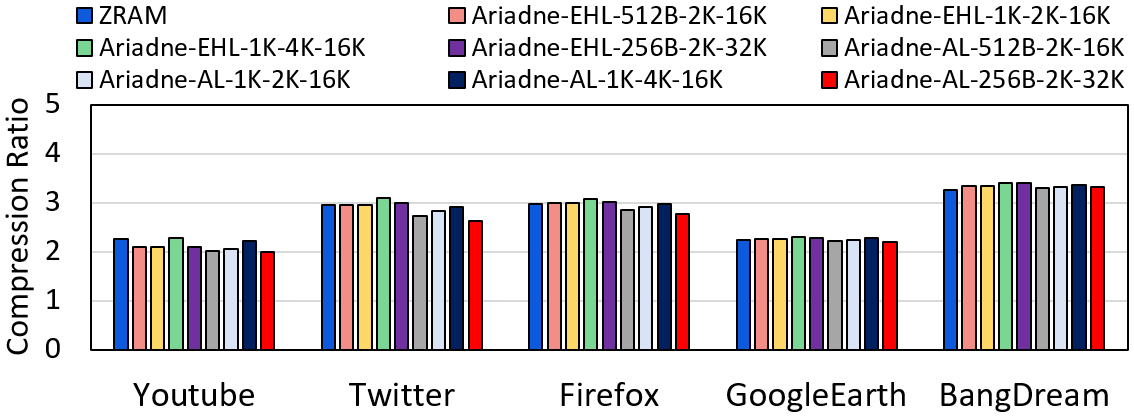}
\caption{Compression ratios under different compressed swap schemes. Higher values are better.}
\label{fig:compression-ratio-result}
\end{figure}

\noindent\textbf{Accuracy and coverage of hot data identification.} Figure~\ref{fig:accuracy} shows the \emph{Coverage} and \emph{Accuracy} of hot data identification for all the evaluated applications. \emph{Coverage} refers to the percentage of correctly predicted data of an application relaunch, and \emph{Accuracy} denotes the percentage of data in the hot list that will be utilized next time, including the data used during both relaunch and execution. 
We make two key observations. 
First, \proposal's \emph{Coverage} for hot data is approximately 70\% on average.
When hot data is mistakenly categorized as warm or cold data, it is compressed in larger sizes, which can lead to longer decompression latencies (see Section~\ref{sec:sensitivity} for more detail).
Second, \emph{Accuracy} of hot data identification is approximately 92\%. This means that our prediction incurs a small penalty for storing all data in the hot list in main memory, as 92\% of the stored hot data will be used in the next application relaunch or execution.

\begin{figure}[!h]
\centering
\includegraphics[width=0.44\textwidth]{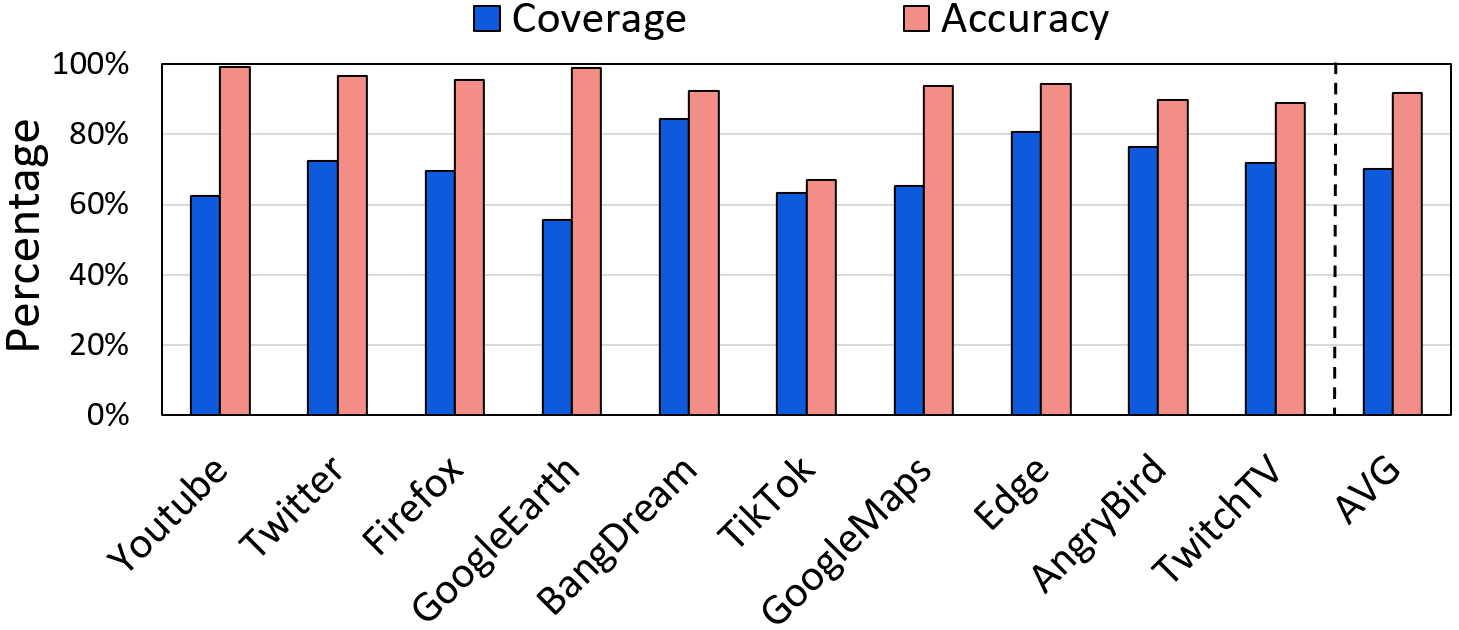}
\caption{Coverage and accuracy of \proposal's hot data identification method for different applications.}
\label{fig:accuracy}
\end{figure}

\subsection{Sensitivity Study}
\label{sec:sensitivity}

We analyze the sensitivity of compression chunk size on compression/decompression latency and compression ratio in \proposal. 
We evaluate two example configurations to illustrate the size configurations' impact on compression and decompression latency as well as compression ratio in Figure~\ref{fig:sensitivity}. The x-axis represents the targeted applications across all three figures. The y-axis, respectively, shows (a) compression latency,  (b) decompression latency, and (c) compression ratio of the data from the targeted application traces. 

\begin{figure}[!h]
\vspace{0.5em}
\centering
\includegraphics[width=0.485\textwidth]{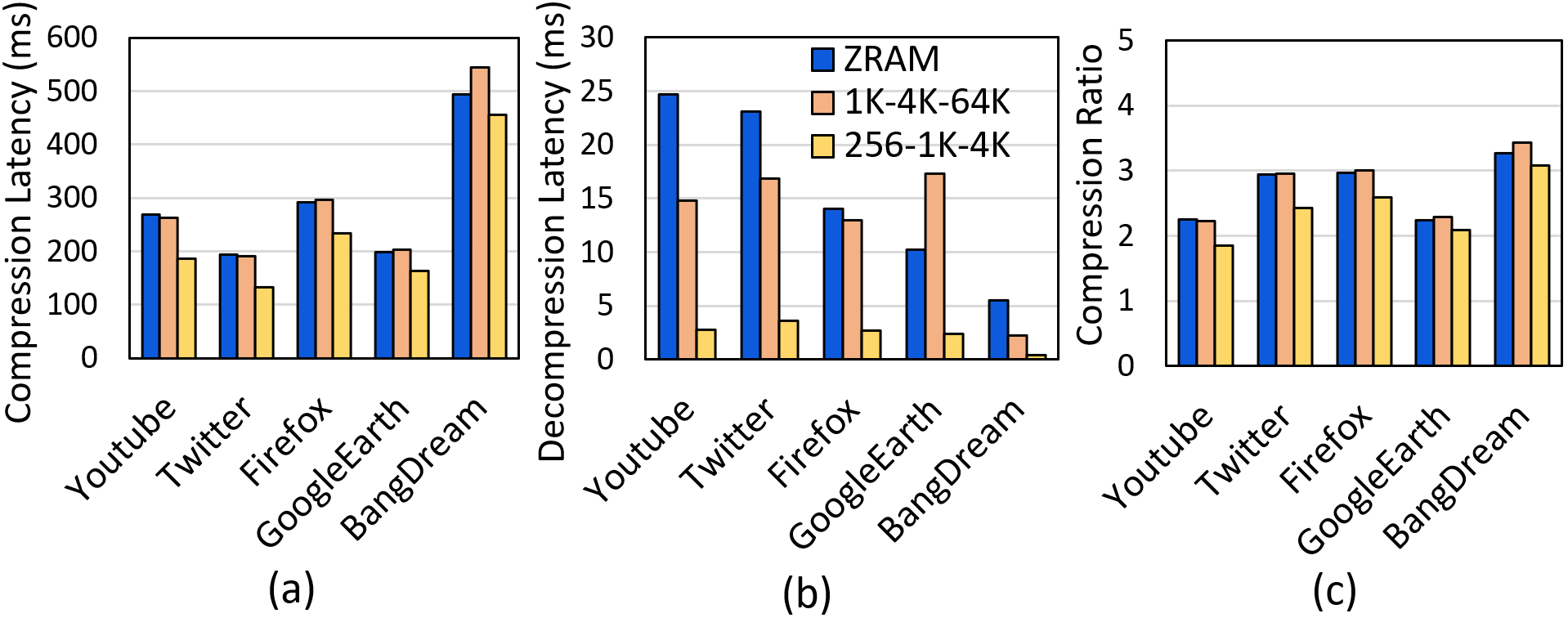}
\caption{Sensitivity study: Compression latency (a), decompression latency (b), and compression ratio (c) under \emph{ZRAM}, \emph{Ariadne-AL-1K-4K-64K}, and \emph{Ariadne-AL-256-1K-4K}.}
\label{fig:sensitivity}
\end{figure}

We make two observations. 
First, selecting inappropriate compression chunk sizes for different hotness levels of data either increases the compression and decompression latencies or reduces the compression ratio. 
Second, using a very large compression chunk size for cold data increases the compression ratio without the penalty of long decompression latency. 
However, it also carries significant risks of potential performance loss if data profiling is inaccurate. If hot or warm data is misclassified as cold data, it gets compressed using a larger chunk size, resulting in longer decompression latencies and worse user experience during application relaunch. Thus, we avoid using excessively large chunk sizes (e.g., $\geq$ 64K) even for cold data. 

\subsection{Overhead Analysis}

We analyze the memory capacity and CPU overhead for all three techniques: \dataorg, \compress, and \predi.
First, \dataorg achieves hotness-aware data organization without physically moving data. Instead, it employs a new data selection policy during compression by operating on LRU lists. This policy does not affect application execution, relaunch latency, or energy consumption, as it only involves operations on the LRU lists,  increasing them slightly over the baseline ZRAM system. 
Specifically, \proposal increases LRU list operations to move part of the previous application's hot data into the warm list when relaunching a new application. 
Since an LRU list operation is much faster (e.g., 100$\times$~\cite{DRAMFAST}) than swapping~\cite{kim2019ezswap,zhu2017smartswap}, the overhead is negligible. 
Second, \compress could introduce memory capacity and CPU overhead if it compresses data used at different times together, as discussed in Section~\ref{sec:compress}. Thus, there is no overhead on hot and warm data, as we use small-size (i.e., smaller than one page) compression for them, ensuring that all the decompressed data will be used together. The overhead on cold data is negligible, as it is unlikely to be accessed again due to the high identification accuracy, as shown in Figure~\ref{fig:accuracy}.
Third, \predi may result in memory capacity overhead and increased energy consumption if the predictions for pre-decompression are inaccurate. To minimize such overheads, we pre-decompress only one page, ensuring high prediction accuracy, as shown in Table~\ref{tab:locality}. 
In summary, \proposal has small overhead in terms of computation and memory space.

\section{Related Work}

To our knowledge, \proposal is the first work that leverages different compression chunk sizes based on the hotness level of the data, while also performing speculative decompression based on data locality characteristics to improve the performance of compressed swap schemes on mobile devices.
We have already compared \proposal extensively with the state-of-the-art \emph{ZRAM} scheme~\cite{zram1} in Section~\ref{sec:evel}. In this section, we discuss related work in two broad categories: flash memory-based swap schemes and emerging NVM-based swap schemes.

\noindent\textbf{Flash memory-based swap schemes}.  
Several prior works~\cite{end2024more, bergman2022znswap, Changlong2020seal, guo2015mars, saxena2010flashvm, kim2017application, zhu2017revisiting, Samsung-Enable-Swap, Xiaomi-Enable-Swap, kim2019analysis} explore using flash memory-based storage as an extension of main memory. While doing so increases memory capacity, it reduces the lifetime of flash memory due to the increased number of writes. Some prior works\cite{guo2015mars,end2024more} aim to reduce writes to flash memory by reducing interference caused by the Android runtime garbage collector on page swapping.  MARS~\cite{guo2015mars} tracks pages that have undergone runtime garbage collection and avoids swap operations on these pages. MARS also employs several flash-aware techniques to accelerate swap operations. 
Fleet~\cite{end2024more} performs runtime garbage collection only on soon-to-be-invalid data of background applications to reduce unnecessary swap operations on long-lifetime foreground application data.
To further reduce swap latency, SmartSwap~\cite{zhu2017smartswap} predicts the most rarely used applications and dynamically swaps these applications' data to flash memory-based swap space ahead of time.
FlashVM~\cite{saxena2010flashvm} modifies the paging system along code paths for allocating, reading, and writing back pages to optimize the use of storage devices for fast swapping. Flash memory-based swap schemes typically focus on minimizing flash writes or accelerating swap operations via data filtering or efficient page write-back mechanisms. In contrast, the key idea of \proposal is to reduce the frequency and latency of compression, decompression, swap-in, and swap-out operations by leveraging different compression chunk sizes based on the hotness level of the data, while also performing speculative decompression based on data locality characteristics. \proposal can be combined with these prior flash memory-based swap schemes.

Several \emph{ZSWAP}-based works~\cite{zswap, han, kim2019ezswap} aim to leverage both main memory compression schemes (e.g., \emph{ZRAM})~\cite{zram1, zram2, new-zram, merge-zram} and flash memory-based swapping space to reduce writes to flash memory. The key idea of \emph{ZSWAP}~\cite{zswap, han, kim2019ezswap} is to initially move pages to \emph{zpool} and subsequently evict them to secondary storage to accommodate newly incoming pages. 
\emph{ZSWAP} has already been incorporated into \proposal (see Section~\ref{sec:mechanisim}).

An optimization of \emph{ZSWAP}, \emph{ezswap}~\cite{kim2019ezswap}, has two key features: 1) compressing both anonymous and file data, and 2) estimating compression ratios to selectively decide which page to compress. The first feature can be combined with \proposal. While the second can improve  \emph{zpool} efficiency, it comes at the cost of additional compression latency and may impact application relaunch latency. Our evaluation shows \emph{ezswap}'s compression ratio estimation overhead accounts for up to 16.7\% of total compression latency. 
\proposal avoids high-overhead compression ratio estimation by using different compression chunk sizes for hot and cold data.

\noindent\textbf{Emerging NVM-based swap schemes.}  
Several previous works~\cite{zhong2014building, kim2015cause, zhong2017building,kim2018comparison, zhu2017smartswap, liu2017non, oliveira2021extending, oliveira2021extending1} investigate how to efficiently enable swap-based NVMs for mobile devices. These works aim to improve swap scheme performance by separating hot and cold data and efficiently exploiting hardware features. For example, CAUSE~\cite{kim2015cause} introduces a hybrid memory architecture for mobile devices that intelligently allocates DRAM or NVM based on the criticality of the data. Two other prior works\cite{kim2018comparison, zhong2017building} utilize an NVM-based swap space for Android devices, leveraging hot and cold data management to efficiently handle swaps between DRAM and NVM. These works do not leverage the tradeoff between compression latency and compression ratio for mobile workloads with varying data hotness or criticality.  \proposal is complementary to these emerging NVM-based swap schemes.

\section{Conclusion}
State-of-the-art compressed swap schemes to manage limited main memory capacity on mobile devices lead to prolonged application relaunch latency and high CPU usage. To address this problem, we introduced a new compressed swap scheme for mobile devices, called \proposal.
\proposal leverages different compression chunk sizes based on the hotness level of the data, while also performing speculative decompression based on data locality characteristics to improve the performance of compressed swap schemes.
We implement and evaluate \proposal on a real commercial smartphone with a cutting-edge Android operating system. Our experimental evaluation results show that \proposal surpasses the state-of-the-art swap scheme in both application relaunch latency and CPU usage.

\section*{Acknowledgments}{
We thank the anonymous reviewers of HPCA 2025 for their encouraging feedback. We thank the SAFARI Research Group members for providing a stimulating intellectual environment and feedback. We acknowledge the generous gifts from our industrial partners, including Google, Huawei, Intel, and Microsoft. This work is supported in part by the Semiconductor Research Corporation (SRC), the ETH Future Computing Laboratory (EFCL), and the AI Chip Center for Emerging Smart Systems (ACCESS).}

\balance
\bibliographystyle{IEEEtran}
\bibliography{refs}

\end{document}